\pdfoutput=1

\documentclass[mathleft
]{an}
\usepackage{graphicx}
\usepackage{times}
\usepackage{natbib}             
\bibpunct{(}{)}{;}{a}{}{,}      
\newcommand{\nm}[1]{_{\mbox{\scriptsize #1}}}
\newcommand{\kms}{\ensuremath{\mathrm{km\,s}^{-1}}}
\newcommand{\cahk}{Ca\,{\sc ii} H \&\ K\ }
\newcommand{\Msun}{\ensuremath{\mathrm{M}_\mathrm{\odot}}}
\newcommand{\type}[2]{#1\,{\sc #2}}
\def\degr{\hbox{$^\circ$}}

\begin{document}

\Pagespan{1}{}
\Yearpublication{}%
\Yearsubmission{}%
\Month{}%
\Volume{}%
\Issue{}%

\title{The chromospherically active binary star EI~Eridani}
\subtitle{II. Long-term Doppler imaging}

\author{A. Washuettl\thanks{Visiting Astronomer, Kitt Peak National Observatory and
      National Solar Observatories, National Optical Astronomy Observatory, which is operated
      by the Association of Universities for Research in Astronomy, Inc. (AURA) under cooperative
      agreement with the National Science Foundation.}
\and K. G. Strassmeier$^\star$
\and M. Weber$^\star$}
\institute{
        Astrophysical Institute Potsdam (AIP), An der Sternwarte 16,
        D-14482 Potsdam, Germany\\
        AWashuettl\,/\,KStrassmeier\,/\,MWeber@aip.de
}
\titlerunning{The active binary star EI~Eridani: Long-term Doppler imaging}
\authorrunning{A. Washuettl et al.}

\received{} \accepted{} \publonline{}


\abstract{%
Data from 11 years of continuous spectroscopic observations of the
active RS~CVn-type binary star EI~Eridani -- gained at
NSO/McMath-Pierce, KPNO/Coud\'e Feed and during the MUSICOS~98
campaign -- were used to obtain 34 Doppler maps in three
spectroscopic lines for 32 epochs, 28 of which are independent of
each other. Various parameters are extracted from our Doppler
maps: average temperature, fractional spottedness, and
longitudinal and latitudinal spot-occurrence functions. We find
that none of these parameters show a distinct variation nor a
correlation with the proposed activity cycle as seen from
photometric long-term observations. This suggests that the
photometric brightness cycle may not necessarily be due to
just a cool spot cycle. The general morphology of the spot
pattern remains persistent over the whole period of 11 years. A
large cap-like polar spot was recovered from all our images. A
high degree of variable activity was noticed near latitudes of
$\approx$60--70\degr\ where the appendages of the polar spot
emerged and dissolved.}

\maketitle

\section{Introduction}
Our Sun exhibits cyclic behaviour which can be seen in its spectral
output as well as in its spot coverage
\citep{willson:mordvinov03}. Apart from the commonly known
11-years Sunspot cycle, our Sun shows an 80 years long so-called
Gleissberg cycle and an 200--300 years long pseudo-cycle (Wolf,
Sp\"orer, Maunder). Even longer cycles or variations are seen,
and there is evidence that periods of cyclic behaviour alternate
with times of constant, very low activity (e.g. Maunder minimum).
Solar-type G and K stars can exhibit the same
solar-like cycles in their photospheric output as originally
discovered from their chromospheric \cahk emission
\citep{baliunas:soon95,gray:baliunas96}. However, a
photosphere-chromosphere correlation cannot be easily reproduced
for evolved G and K stars \citep{choi:soon95,frasca:biazzo05}.
Long-term changes in the mean brightness of RS~CVn stars are
therefore not a straightforward tool for investigating spot
cycles similar to the Sun's 11-year cycle.
%
From what we know from our Sun, we would expect an activity
cycle from \cahk to be accompanied by a spot cycle.

Several groups began long-term
Doppler-imaging studies on a few selected active stars (that
usually consist of one or a few images per year, i.e. per
observing season): \citet[][ FK~Com, 24 maps from 1993 to
2003]{heidi07}, \citet[][ HR~1099\,=\,V711~Tau, 23 maps from 1981
to 1992]{vogt:hatzes99}, \citeauthor{berdy:etal98c}
(\citeyear{berdy:etal98c}, II~Peg, 9 maps from 1992 to 1996; and
\citeyear{berdy:etal01}, LQ~Hya, 9 maps from 1993 to 1999),
\citet[][ UZ~Lib, 8 maps from 1994 to 2000]{olah:kgs:weber02uzlib},
\citet[][ IM~Peg, 31 maps from 2003 to 2005]{marsden:berdy07},
\citet[][ AB~Dor, LQ~Hya, HR\,1099]{donati:acc03}
and others. All of them are active stars and most of them are
close binaries with their rotation period usually spun up by tidal
forces. However, so far only one star, HR\,1099
\citep{vogt:hatzes96,paperxii}, has been observed over a sufficiently
long time span to cover a complete activity cycle with Doppler maps.

EI~Eridani = HD 26337 (G5\,{\sc iv}, $P\nm{rot}\ = 1.947$ days,
$V$ = 7.1 mag, $v\sin i = 51$ \kms, SB1) was one of the first
stars being Doppler imaged. Its
spot distribution has been monitored since 1984. Long-term
photometric observations revealed large brightness variations
which seem to be cyclic and were first supposed to resemble a
solar-like 11 years cycle but appear to be more complex as
observations continue. Within the first 16 years of photospheric
observations, EI\,Eri gave the impression of a cycle of similar
length to the solar cycle: \citet{kgs:bartus97} calculated a
length of $11 \pm 1$ years. Observations in the following years
did not confirm this cycle as they did not meet the anticipated
decline of brightness. \citet{berdy:etal98b} estimated a 9-year
periodicity from the positions of active regions. They did not
discuss a possible change in the level of spottedness itself,
though. \citet{olah:kollath00} favoured a length of 16.2 years on a
time base of 18.5 years and quote a significant remaining period
of 2.4 years. Adding four more years of data, the preferred cycle
length was back to 12.2 years \citep[see][]{olah:kgs02}.
Most recent calculations, comprising an observed time interval of
28 years, state a long cycle of about 14 years with variable
amplitude and two short cycles of about 2.9--3.1 years and 4.1--4.9
years \citep{olah:cycle09}.

The present program intends to review all available Doppler maps
of EI\,Eri and to possibly link any long-term evolution of the
spot distribution to the photometric variations.

\section{Doppler imaging of EI\,Eridani}
With its large rotational velocity and an intermediate inclination, EI\,Eri is, as
\citet{fekel:quigley87} already noted, an ideal candidate for {\em Doppler imaging\/} (hereafter ``DI'').
Consequently, EI\,Eri has been a prime target since the first application of this
technique to spotted late-type stars in 1982.
Doppler maps can be found in \citet[][ epoch 1987]{kgs90}, \citet[][ epoch 1988;
investigation of different DI techniques]{kgs:rice91}, and for the 1984-87 period, in \citet{hatzes:vogt92}.
\citet{oneal:saar96} determined spot covering factors between 16\% and 37\%.

The data used for producing the maps presented in this paper were obtained at NSO McMath-Pierce
during seven years of our long term synoptic observations (1988--1995; 19 maps) and one dedicated
visitor observing run in 1996 (three independent maps), several dedicated KPNO/Coud\'e feed
visitor observing runs in 1995, 1996 and 1997 (the latter with three independent maps) and participation in the
MUSICOS~1998 observing campaign (three independent maps). The total covered time period streches from
1988 to 1998 (eleven observing seasons).
A detailed description of the NSO and KPNO observations and the obtained data was given in
\citet[][ hereafter paper I]{paper1}.
The observations and data from the MUSICOS\,98 campaign will be 
presented in a forthcoming paper \citep{paper3} which will also address 
the question of differential rotation.
Doppler imaging suggests abundances of $-6.3$\,dex (Ca) and $-5.5$\,dex which are smaller than solar
by  $-0.6$\,dex (Ca) and $-1.1$\,dex (see below).

Knowledge of the precise stellar parameters is essential for obtaining good-quality
Doppler maps. Parameters like temperature, rotational velocity, luminosity, radius,
orbital inclination, mass, period and orbital parameters were already investigated in paper I.
Additionally, several specific Doppler-imaging parameters were improved by producing
a series of test reconstructions: stellar inclination, differential rotation, rotational
velocity, gravity, temperature, abundances, macro and micro turbulence, line blends
and transition probabilities. Nomalized $\chi^2$ distributions for these parameters
which allow to find or improve the parameter value were presented in \citet{phdwasi}. 
For a list of DI-relevant parameters, see Table~\ref{tab:DIparam}. Further astrophysical
parameters are listed in paper I, Table~5.

\begin{figure}
\begin{center}
\includegraphics[viewport=24 34 515 405,width=82mm,clip,angle=0]{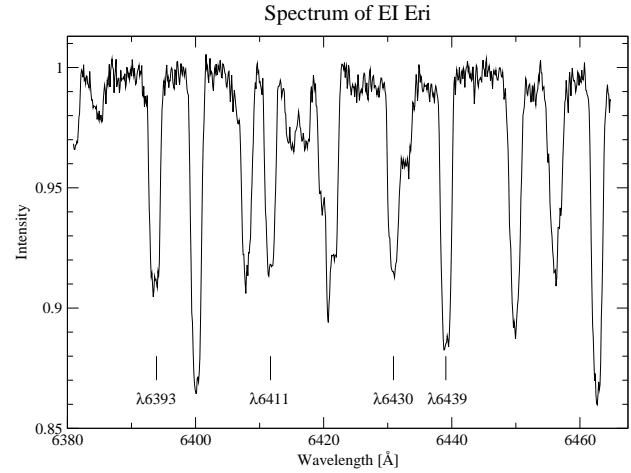}
\caption[Typical spectrum of EI Eri]{Typical spectrum of EI Eri
with the four mapping lines $\lambda$6393, $\lambda$6411, $\lambda$6430 
and $\lambda$6439.}
\label{typical_spectrum}
\end{center}
\end{figure}

\begin{figure*}
\begin{center}
\includegraphics[viewport=38 501 538 723,width=50mm,clip,angle=0]{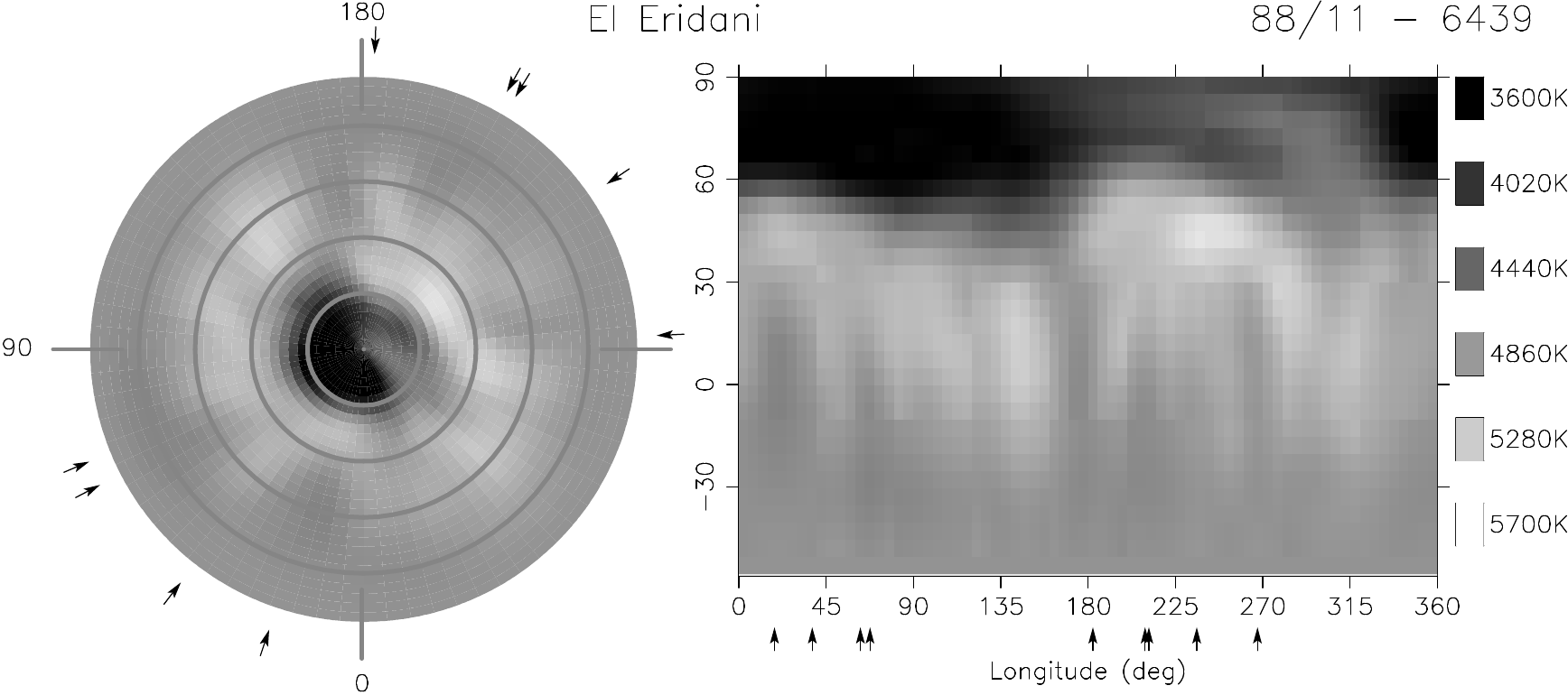}
\includegraphics[viewport=38 501 538 723,width=50mm,clip,angle=0]{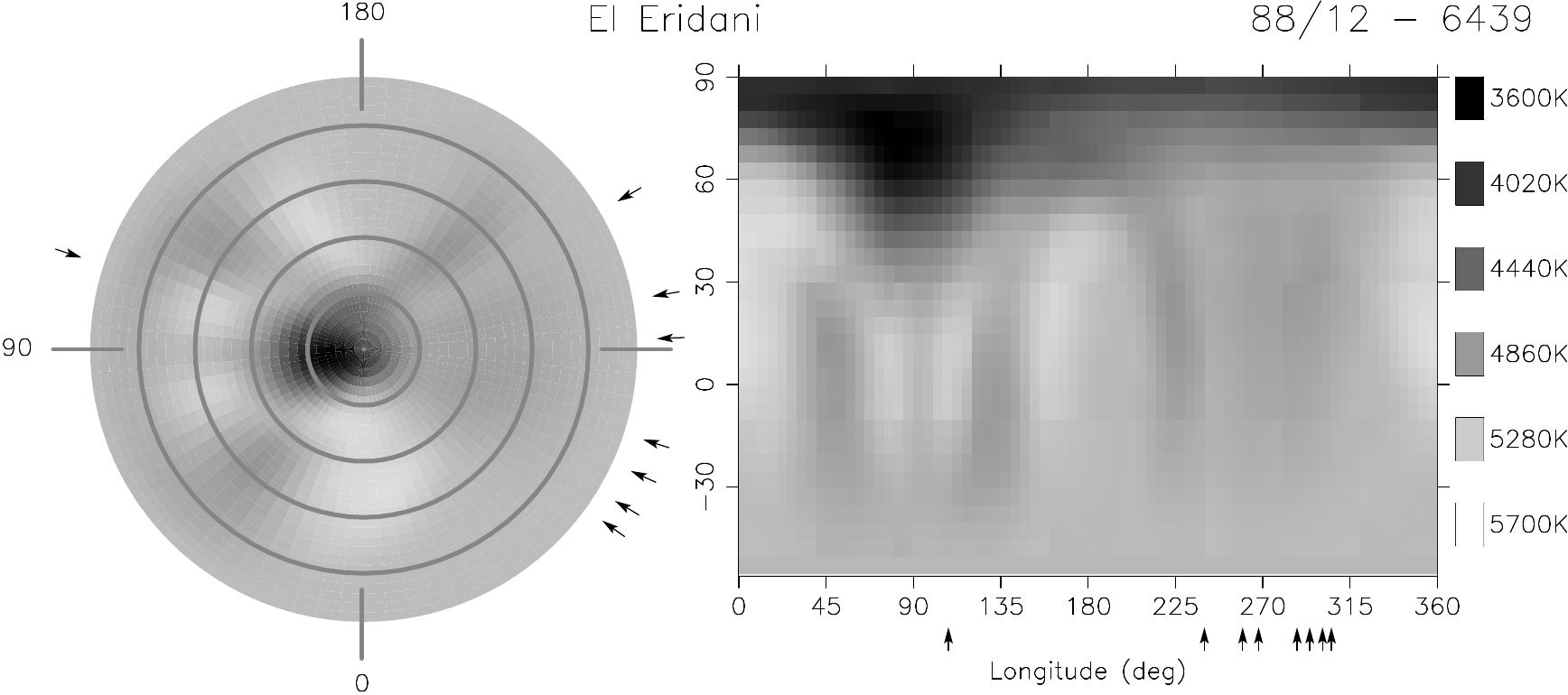}
\includegraphics[viewport=38 501 538 723,width=50mm,clip,angle=0]{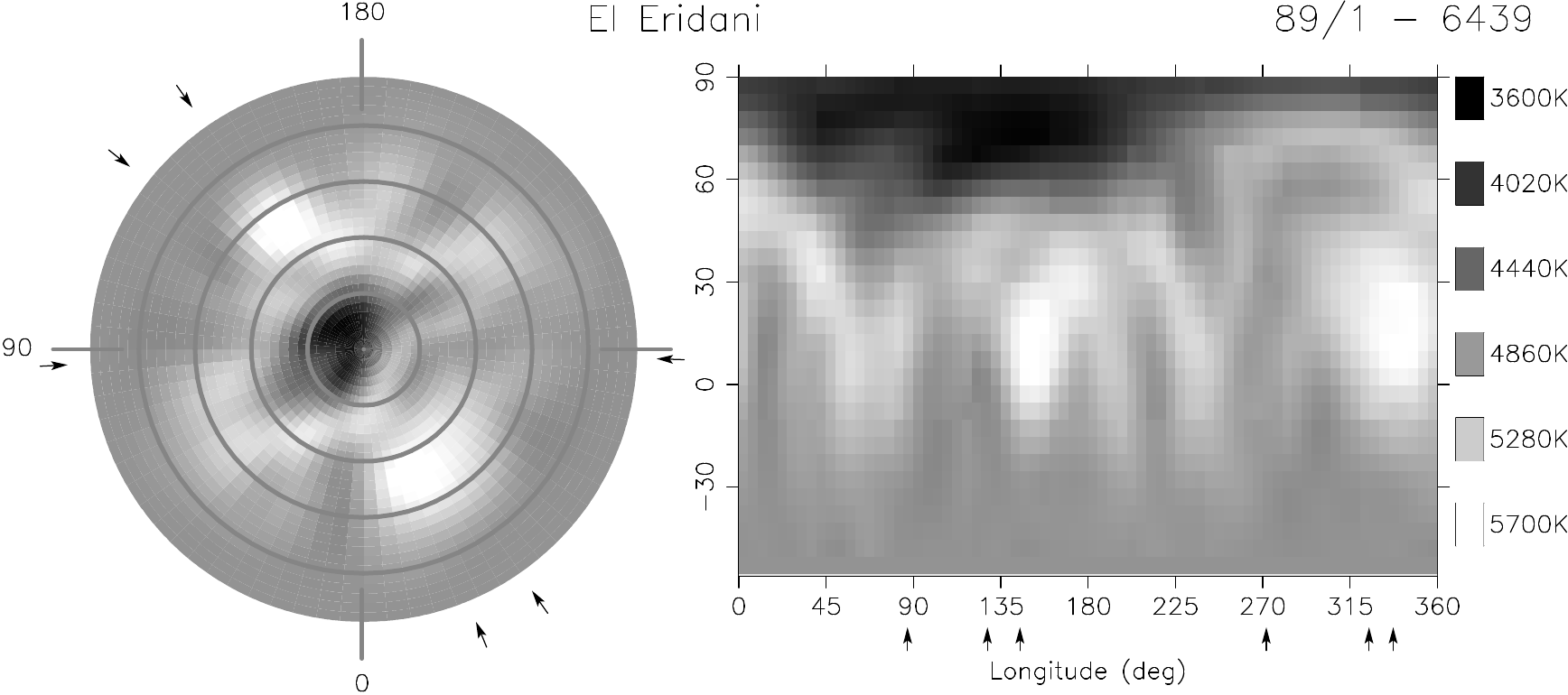}
\includegraphics[viewport=38 501 538 723,width=50mm,clip,angle=0]{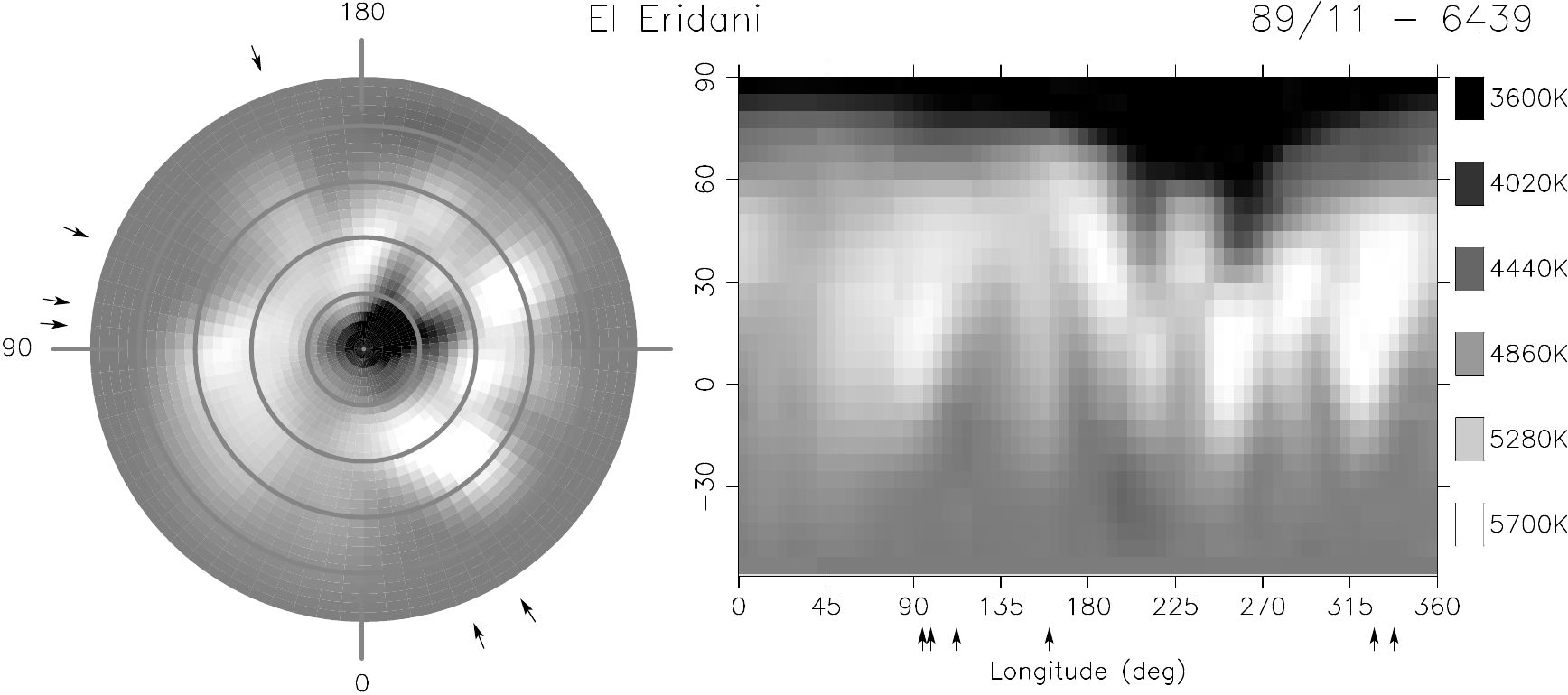}
\includegraphics[viewport=38 501 538 723,width=50mm,clip,angle=0]{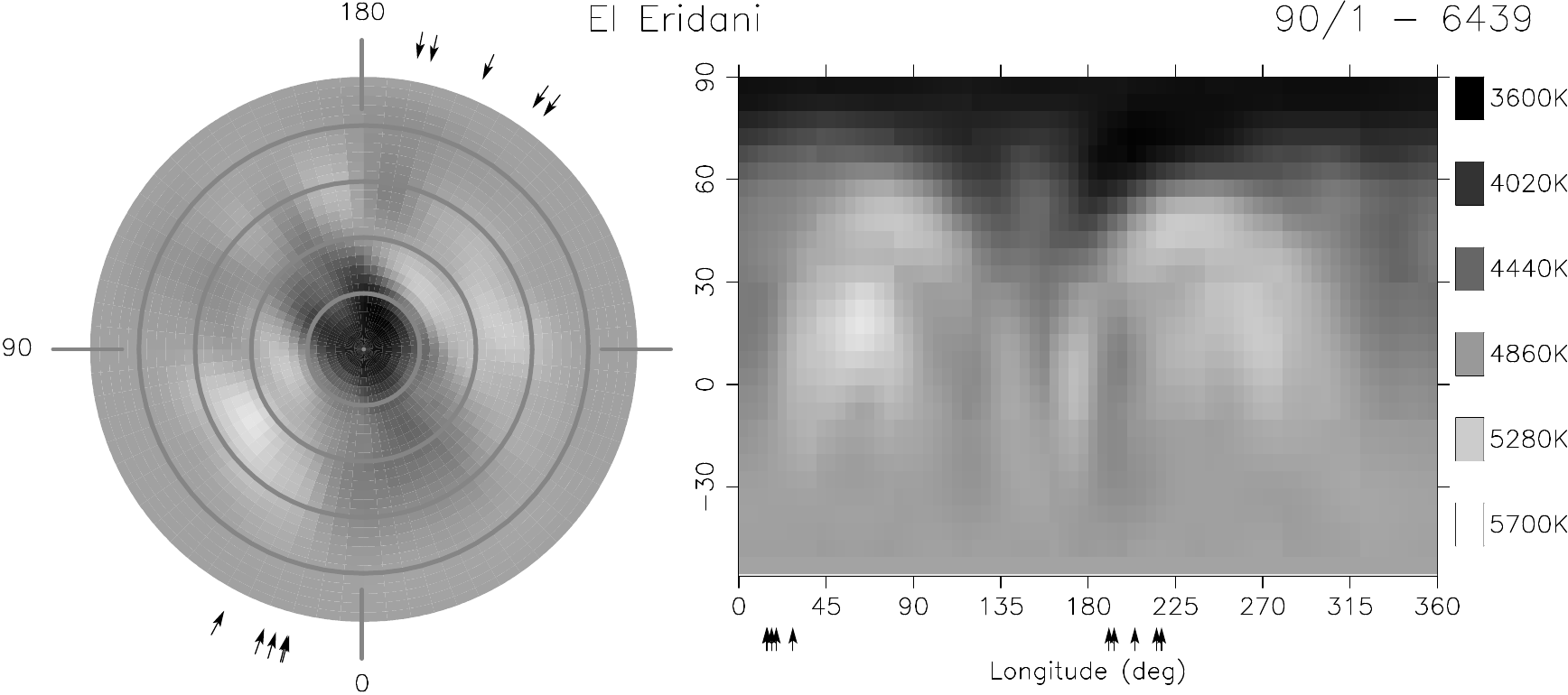}
\includegraphics[viewport=38 501 538 723,width=50mm,clip,angle=0]{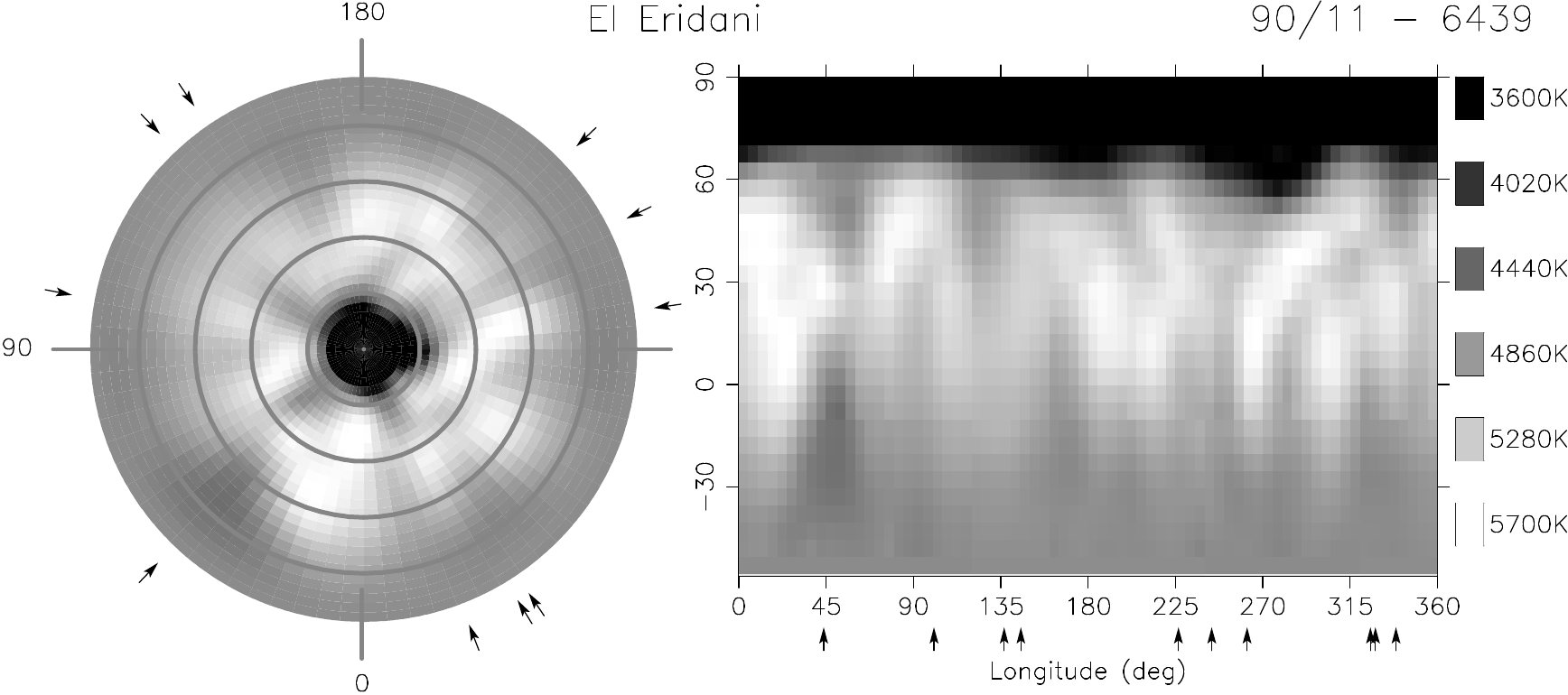}
\includegraphics[viewport=38 501 538 723,width=50mm,clip,angle=0]{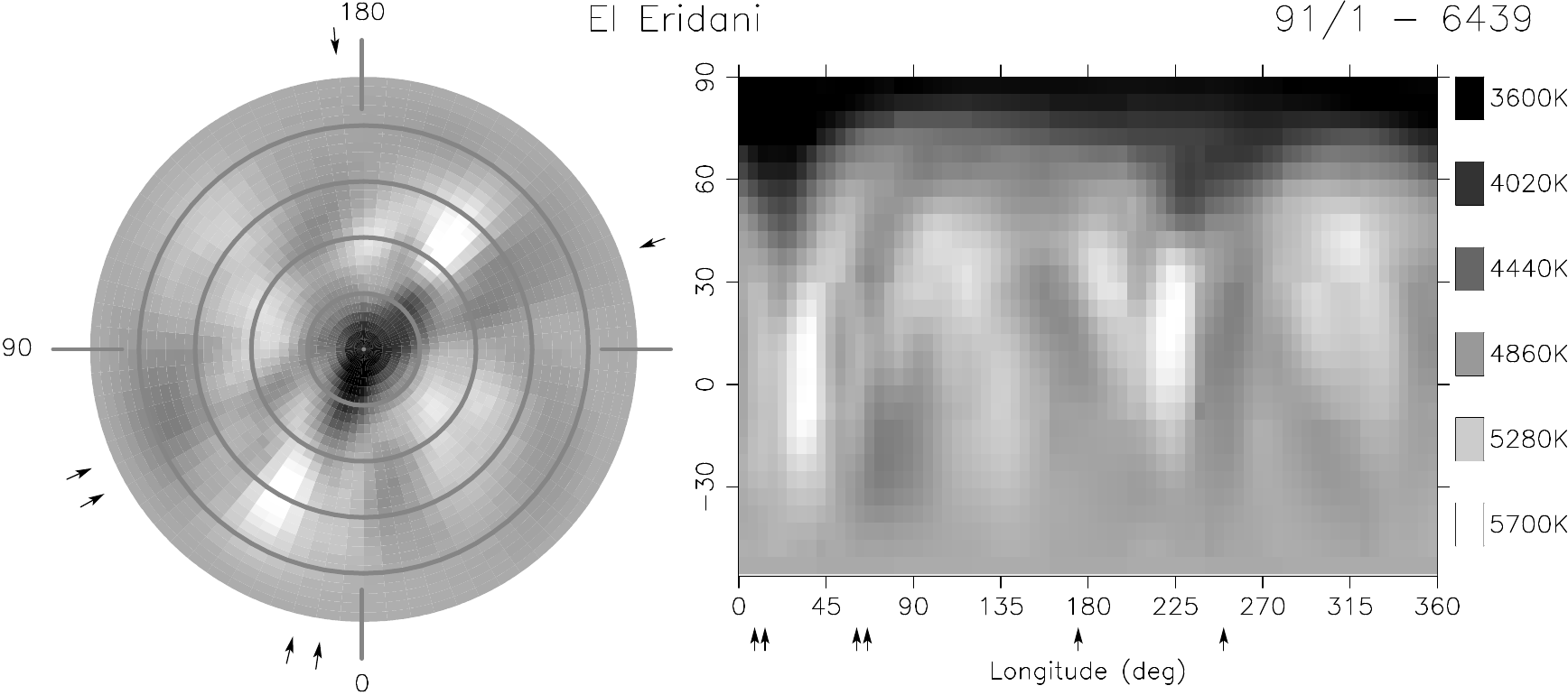}
\includegraphics[viewport=38 501 538 723,width=50mm,clip,angle=0]{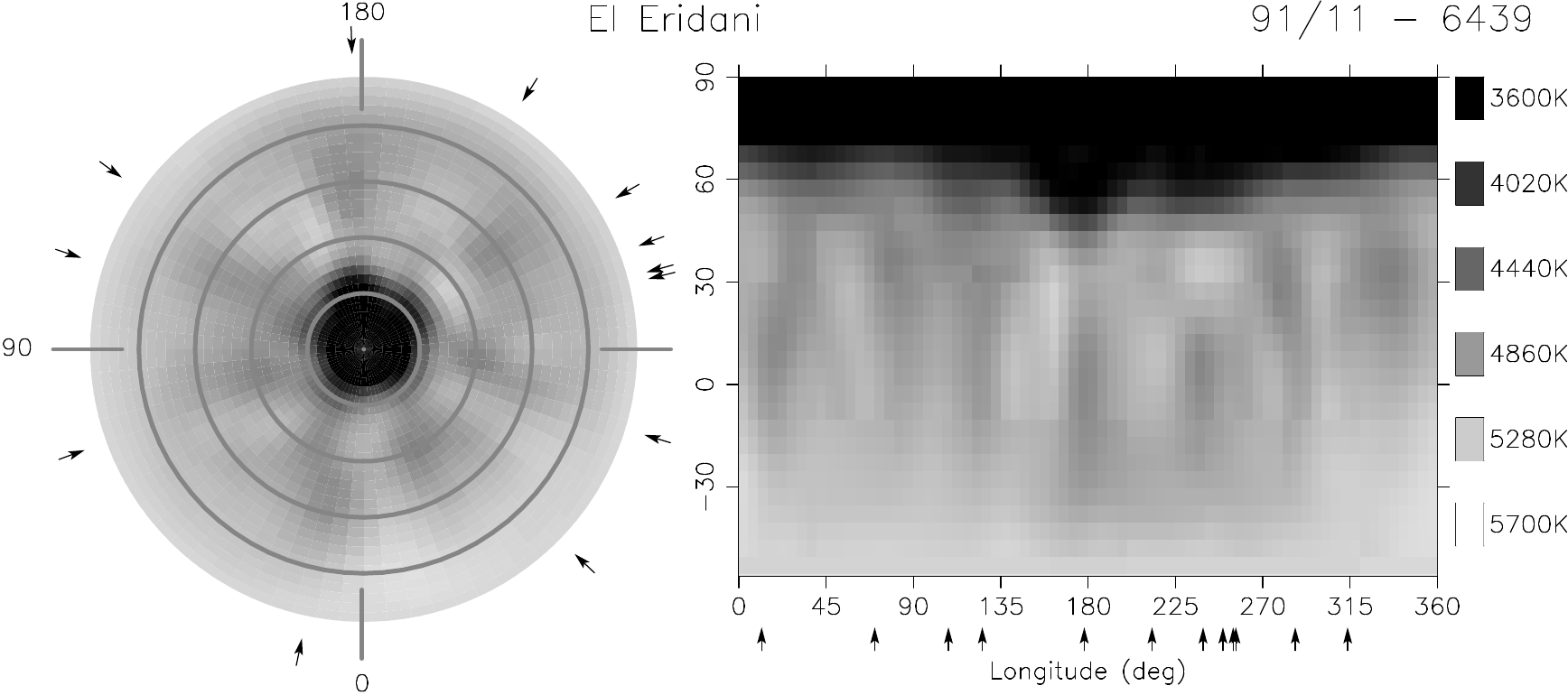}
\includegraphics[viewport=38 501 538 723,width=50mm,clip,angle=0]{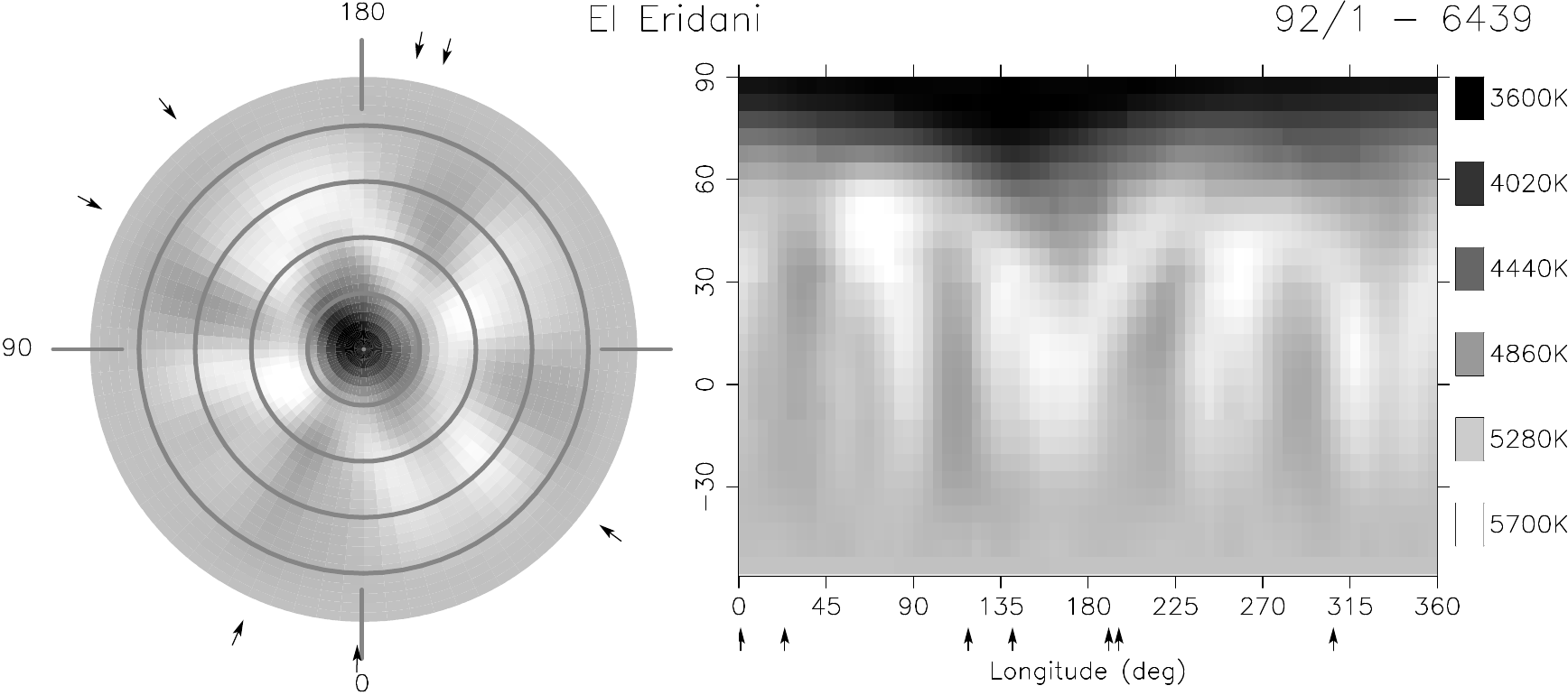}
\includegraphics[viewport=38 501 538 723,width=50mm,clip,angle=0]{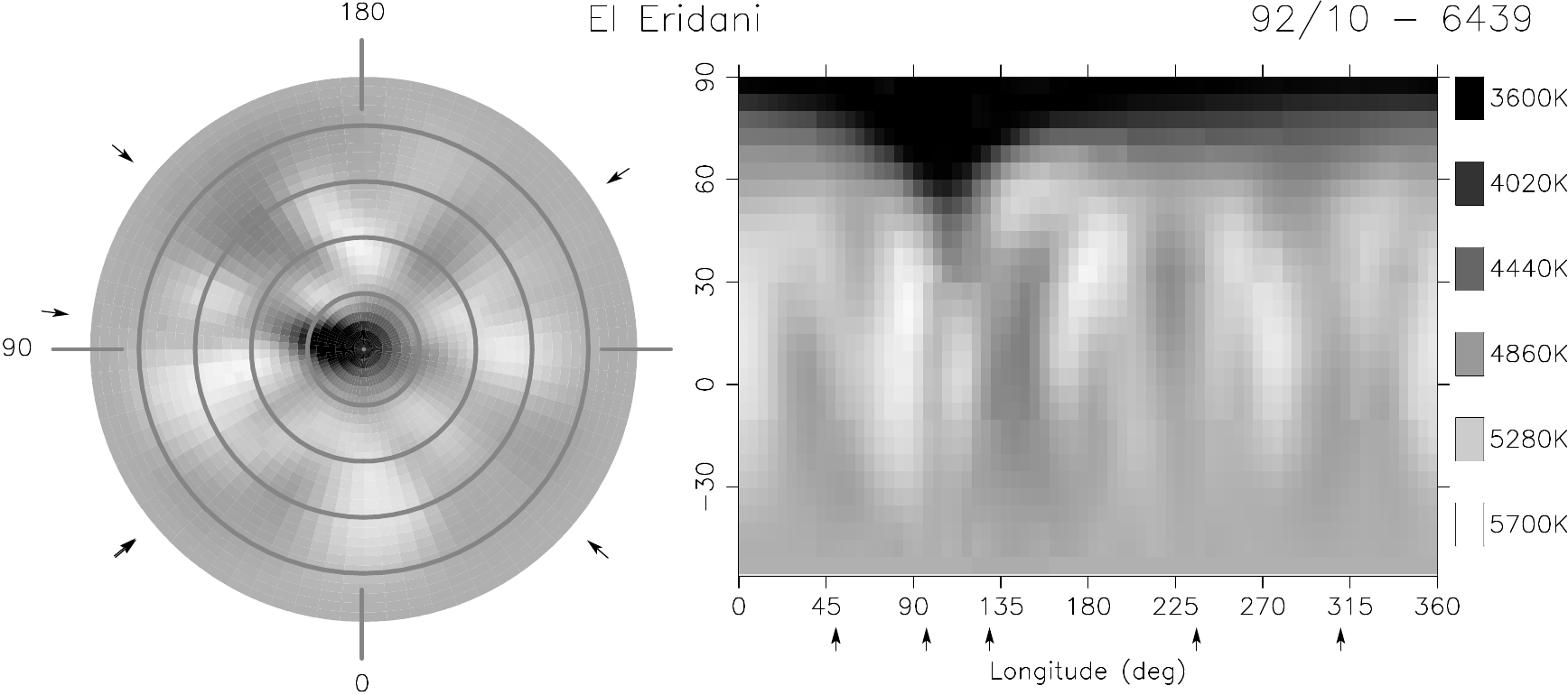}
\includegraphics[viewport=38 501 538 723,width=50mm,clip,angle=0]{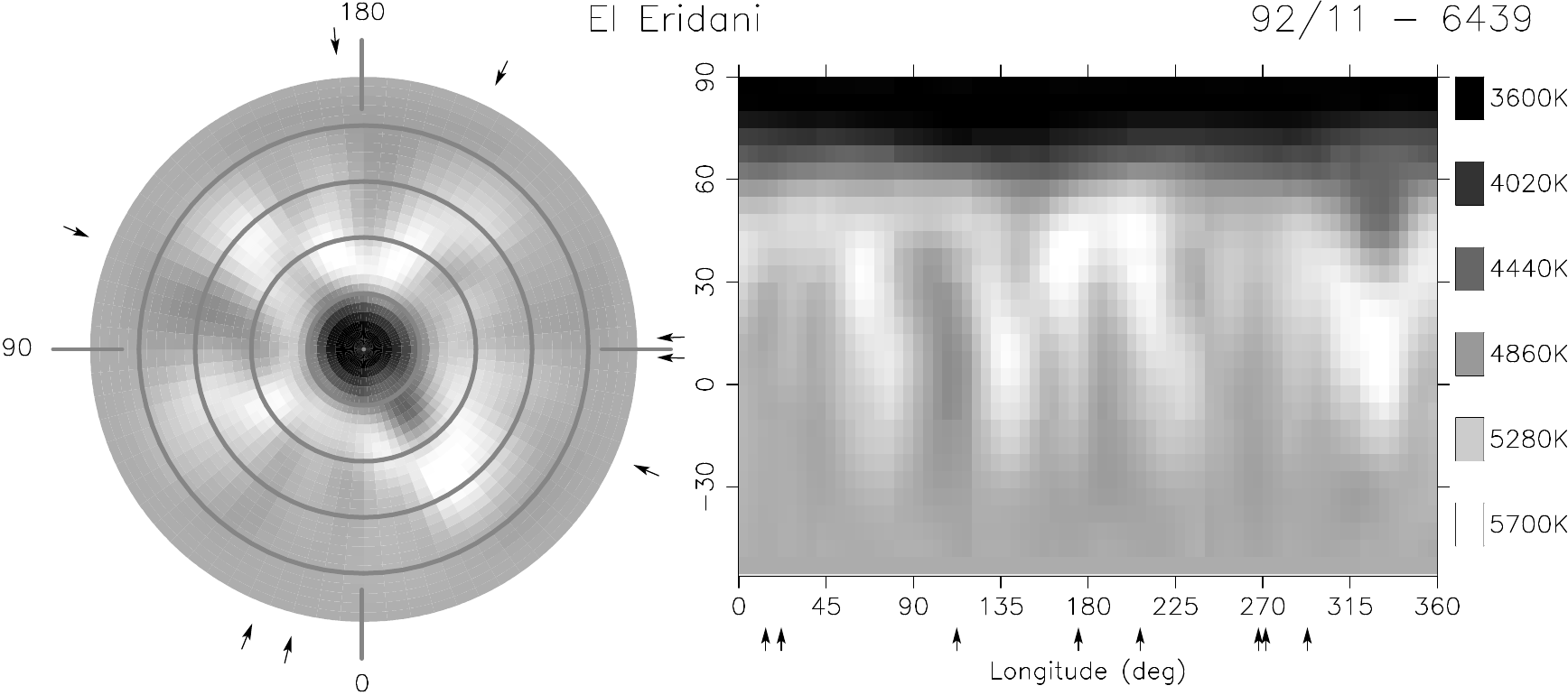}
\includegraphics[viewport=38 501 538 723,width=50mm,clip,angle=0]{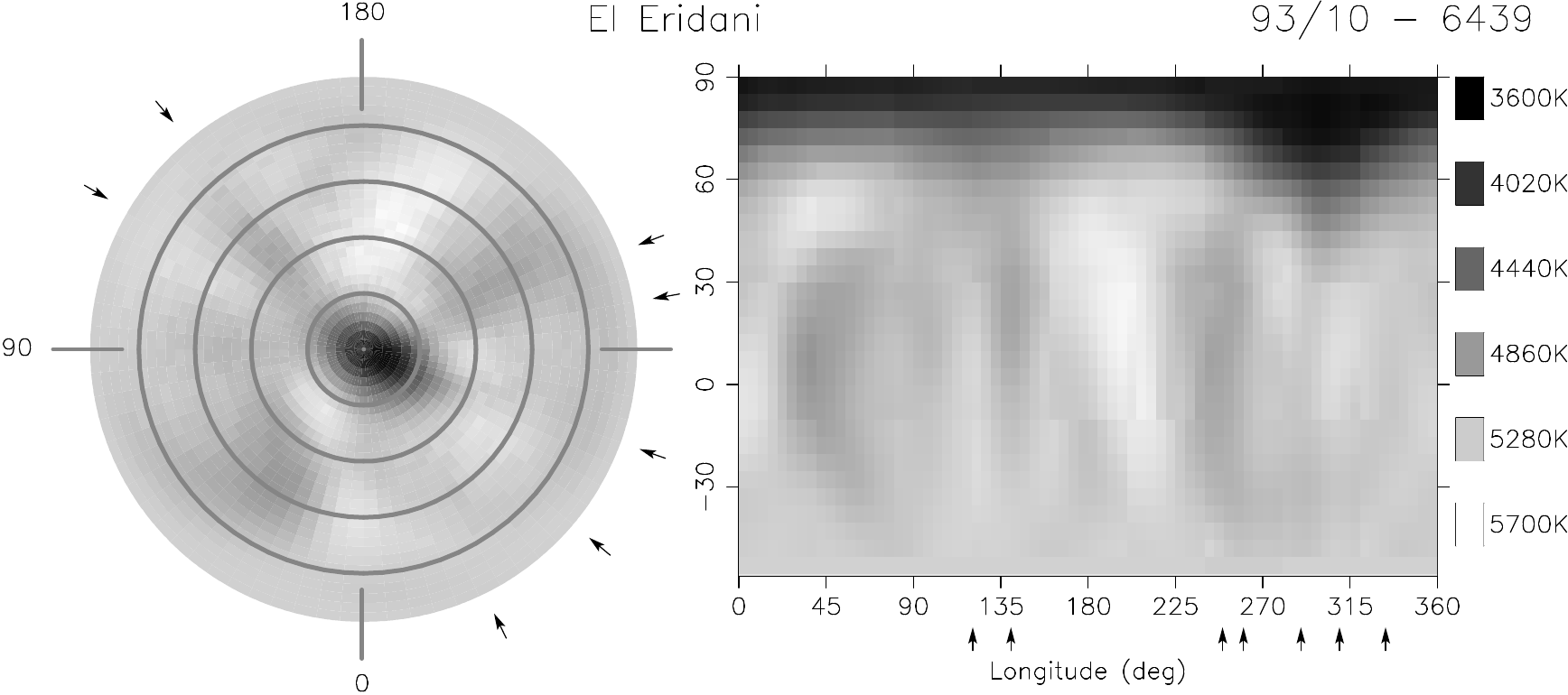}
\includegraphics[viewport=38 501 538 723,width=50mm,clip,angle=0]{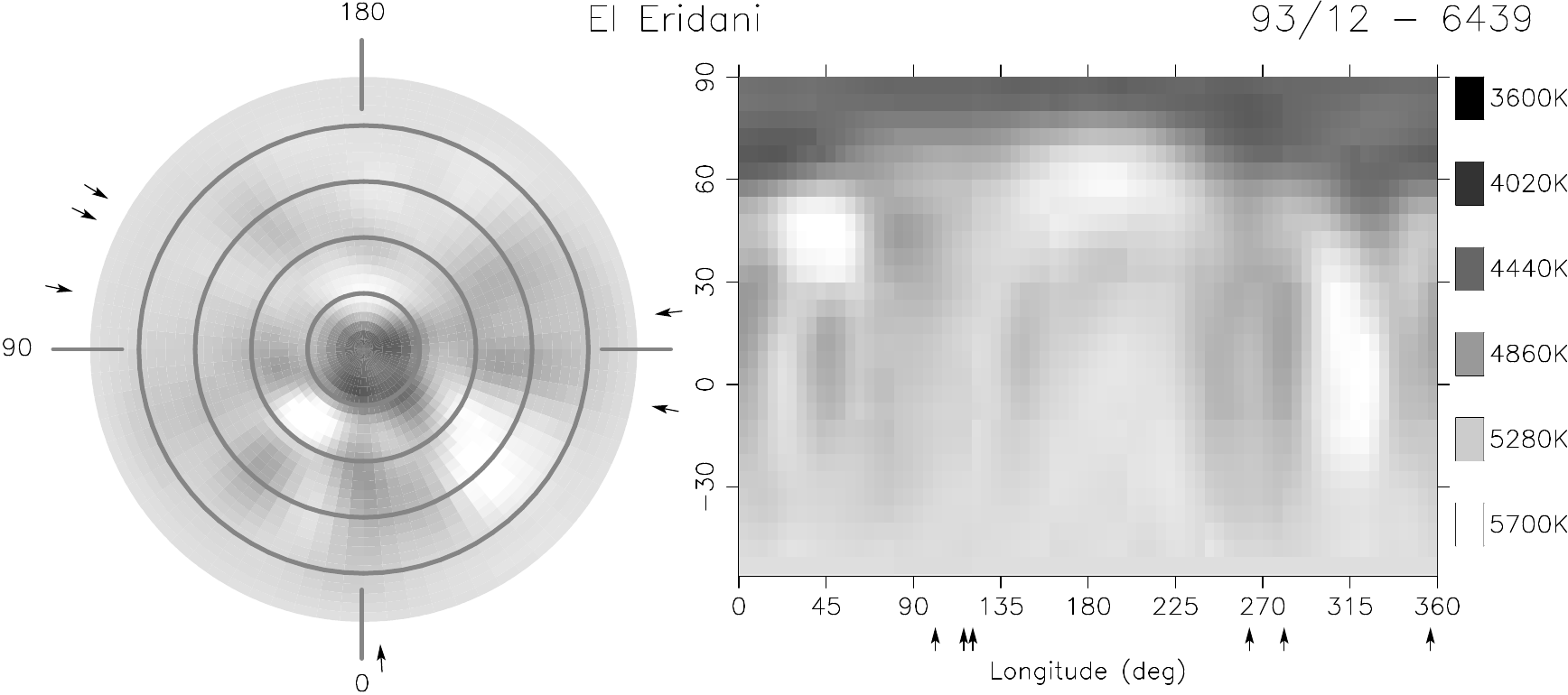}
\includegraphics[viewport=38 501 538 723,width=50mm,clip,angle=0]{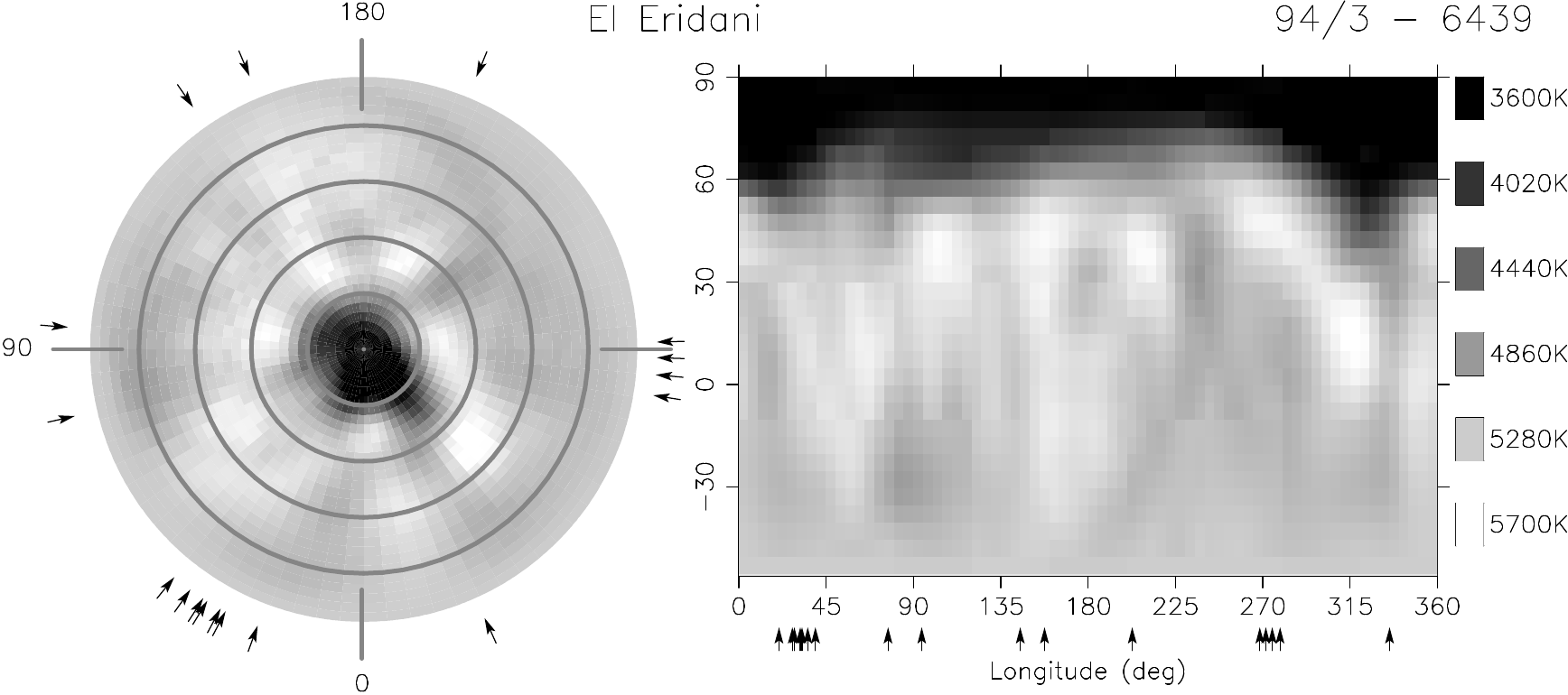}
\includegraphics[viewport=38 501 538 723,width=50mm,clip,angle=0]{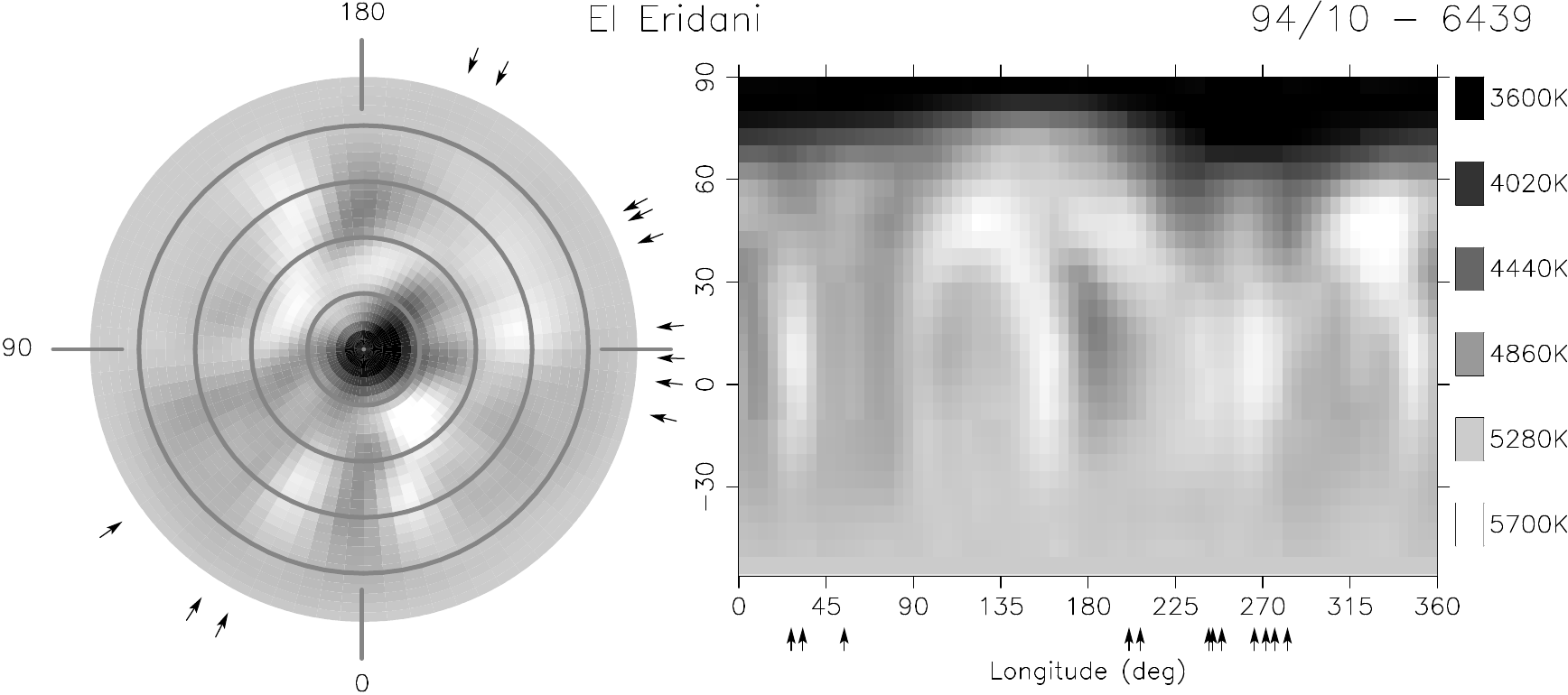}
\includegraphics[viewport=38 501 538 723,width=50mm,clip,angle=0]{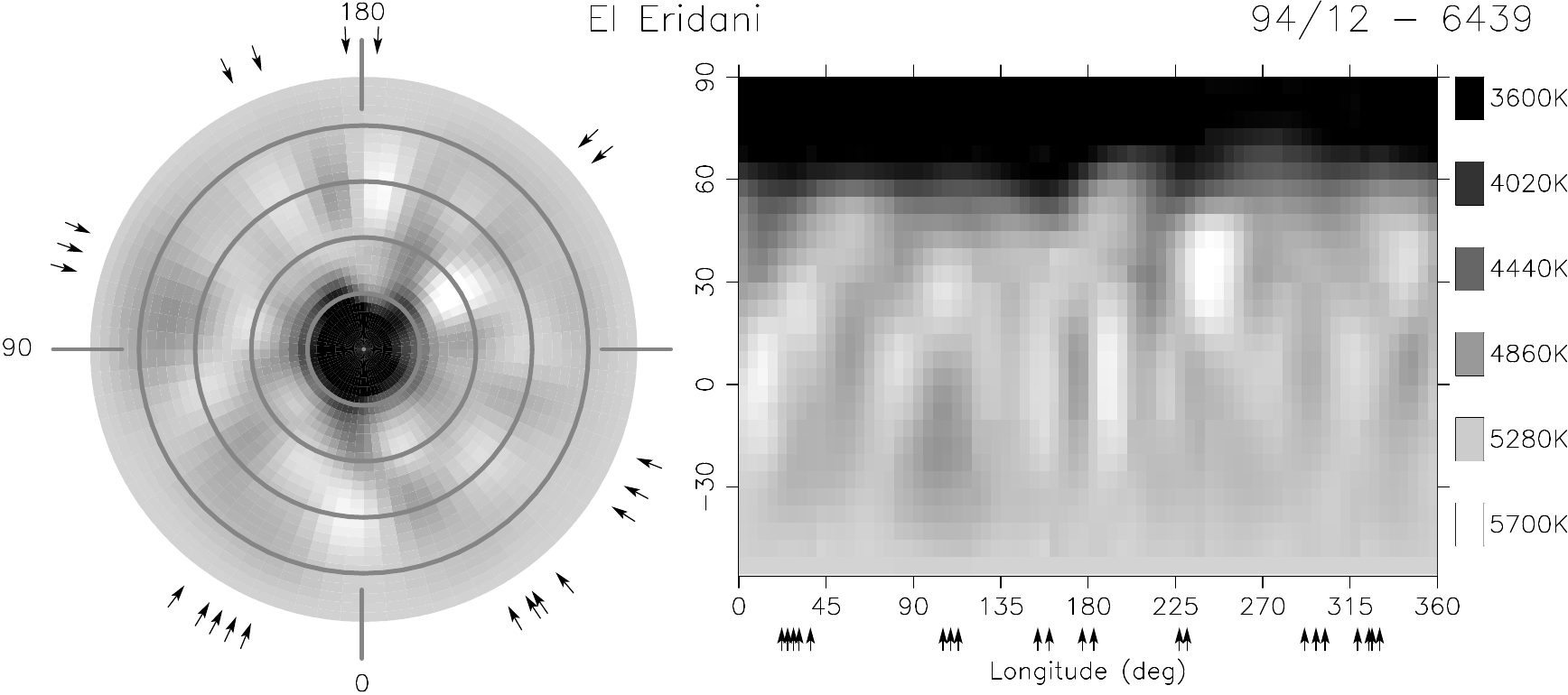}
\includegraphics[viewport=38 501 538 723,width=50mm,clip,angle=0]{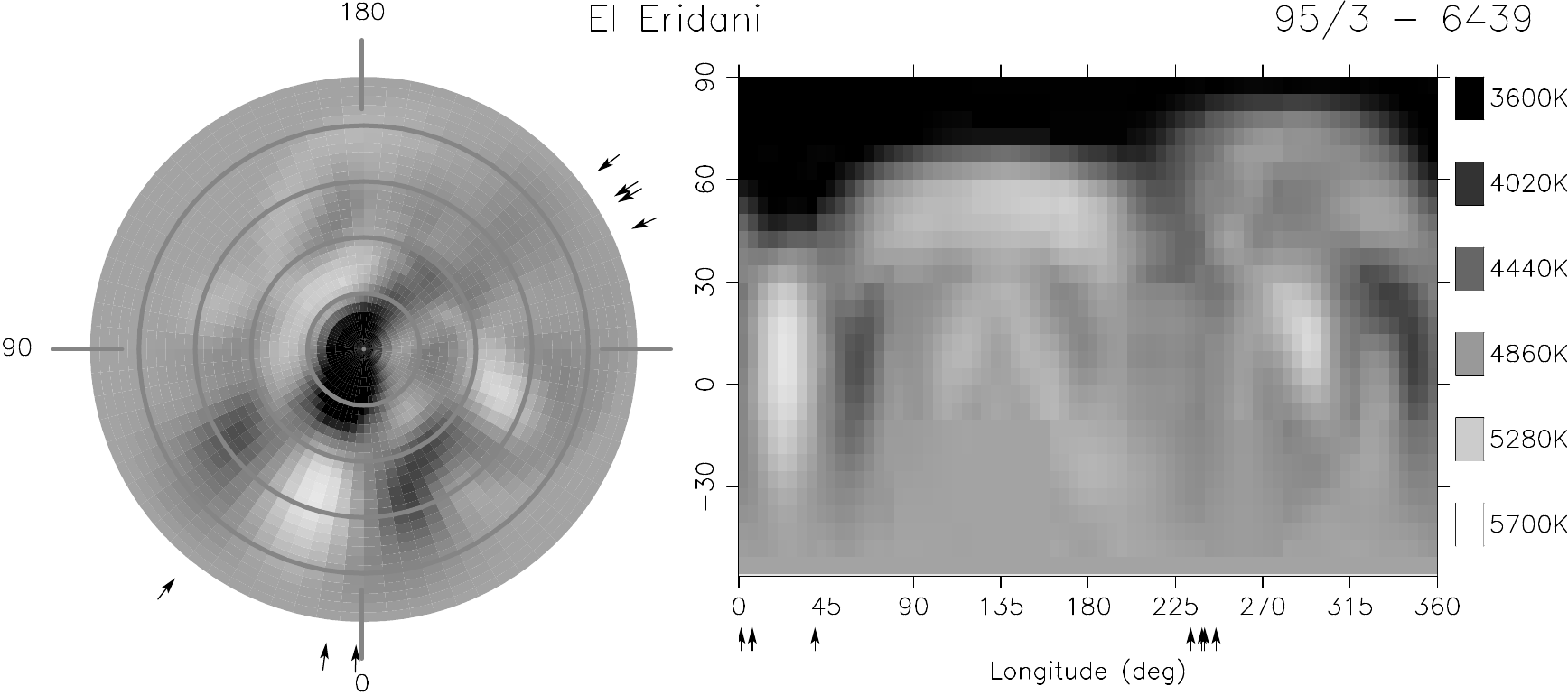}
\includegraphics[viewport=38 501 538 723,width=50mm,clip,angle=0]{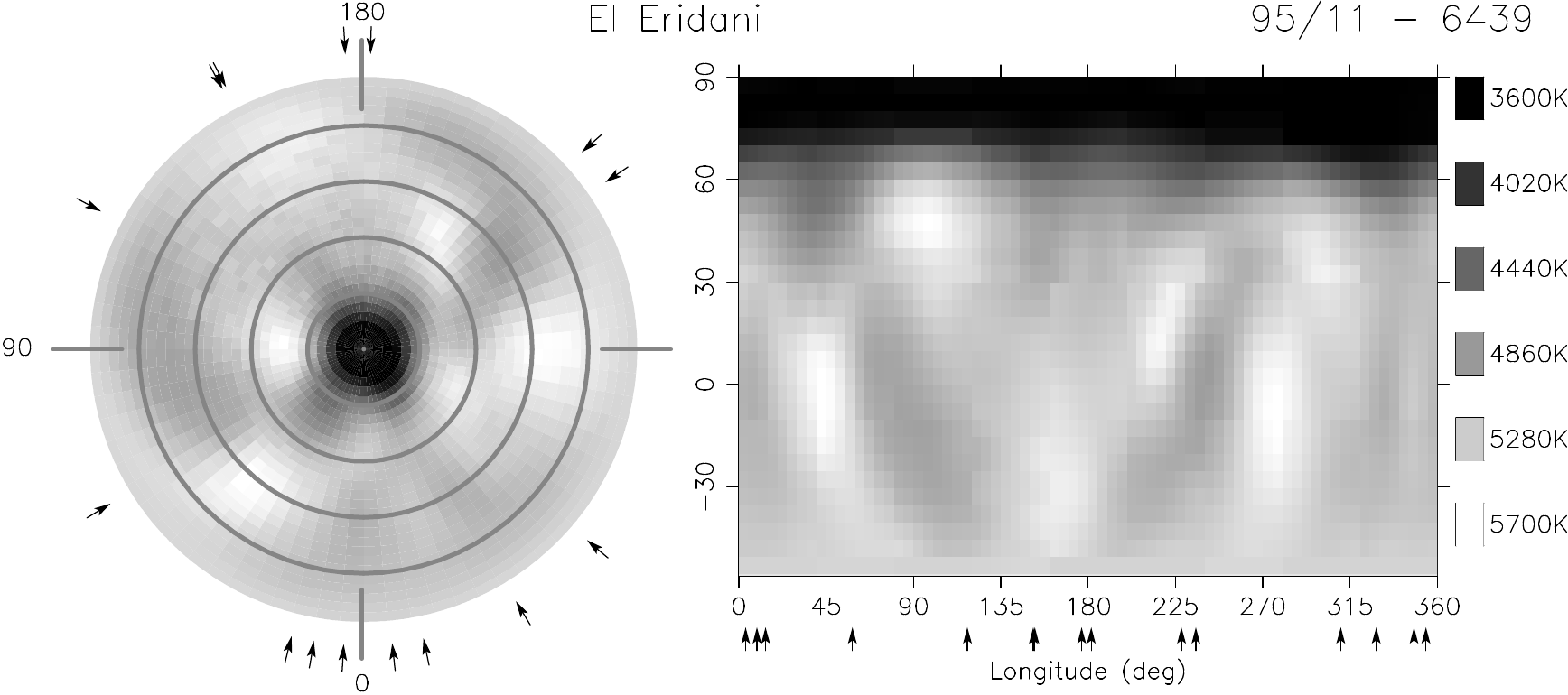}
\includegraphics[viewport=38 501 538 723,width=50mm,clip,angle=0]{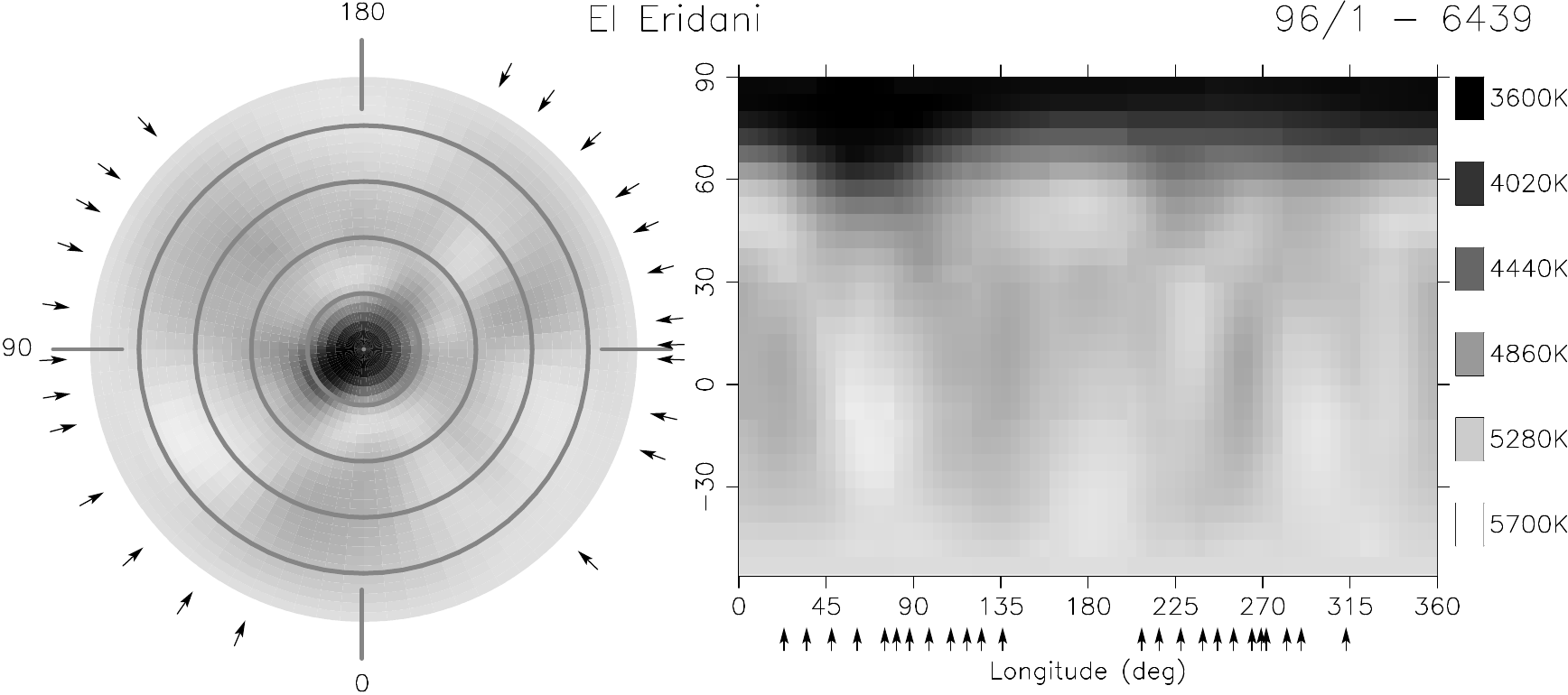}
\includegraphics[viewport=38 501 538 723,width=50mm,clip,angle=0]{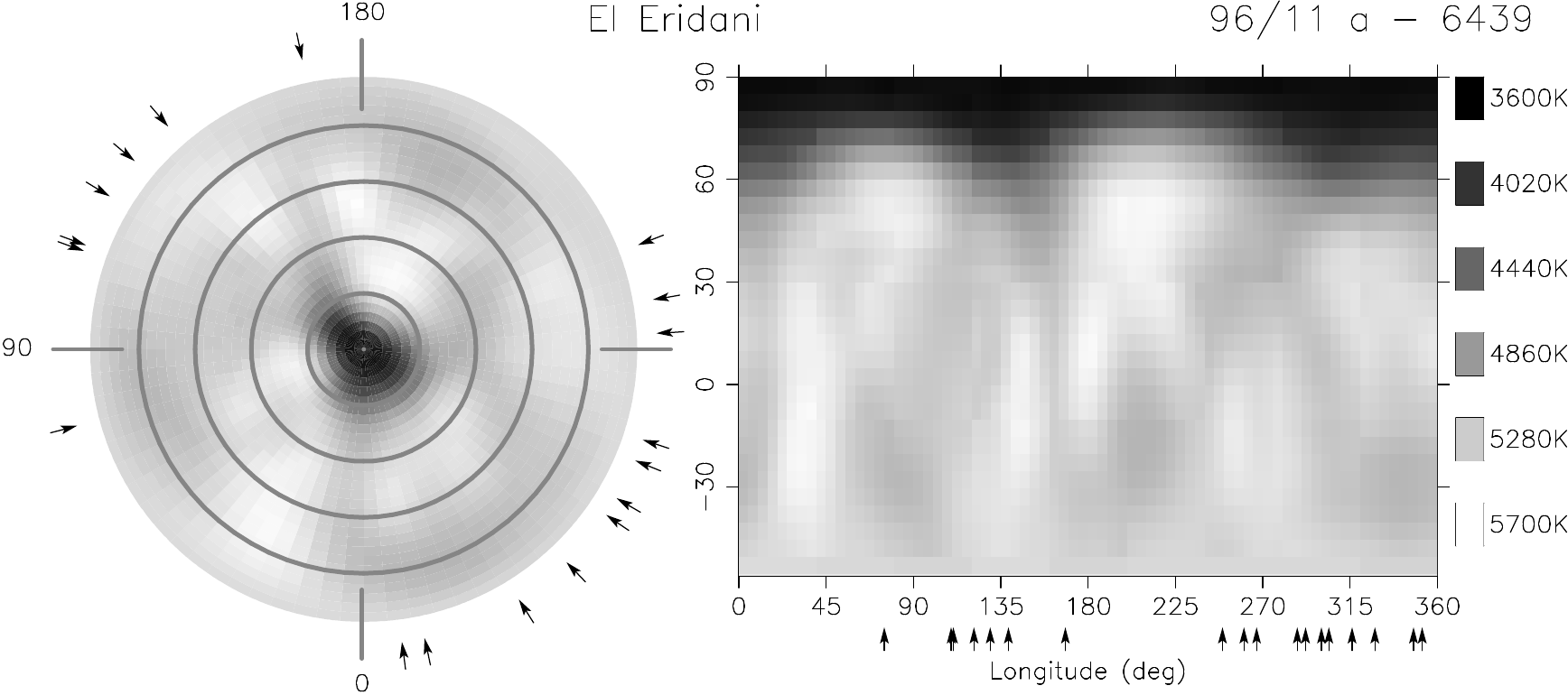}
\includegraphics[viewport=38 501 538 723,width=50mm,clip,angle=0]{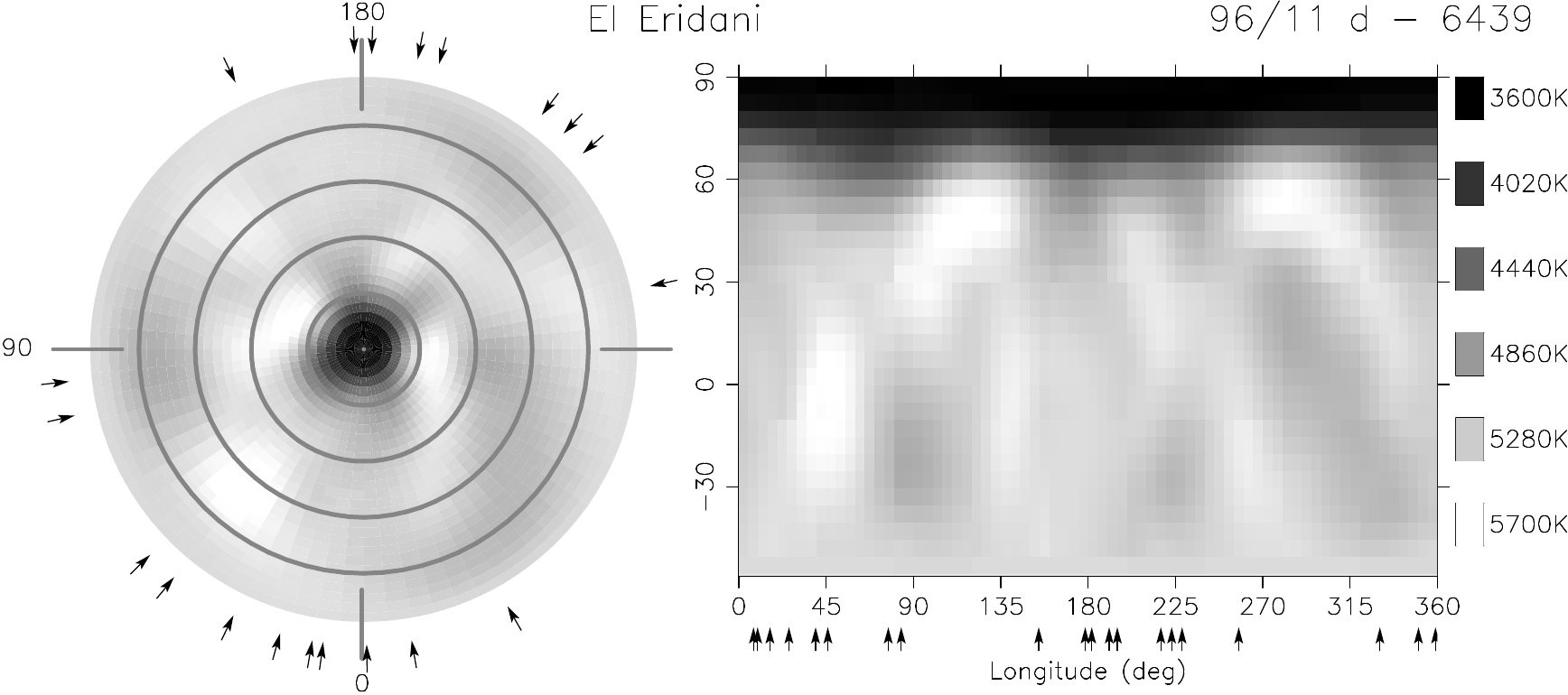}
\includegraphics[viewport=38 501 538 723,width=50mm,clip,angle=0]{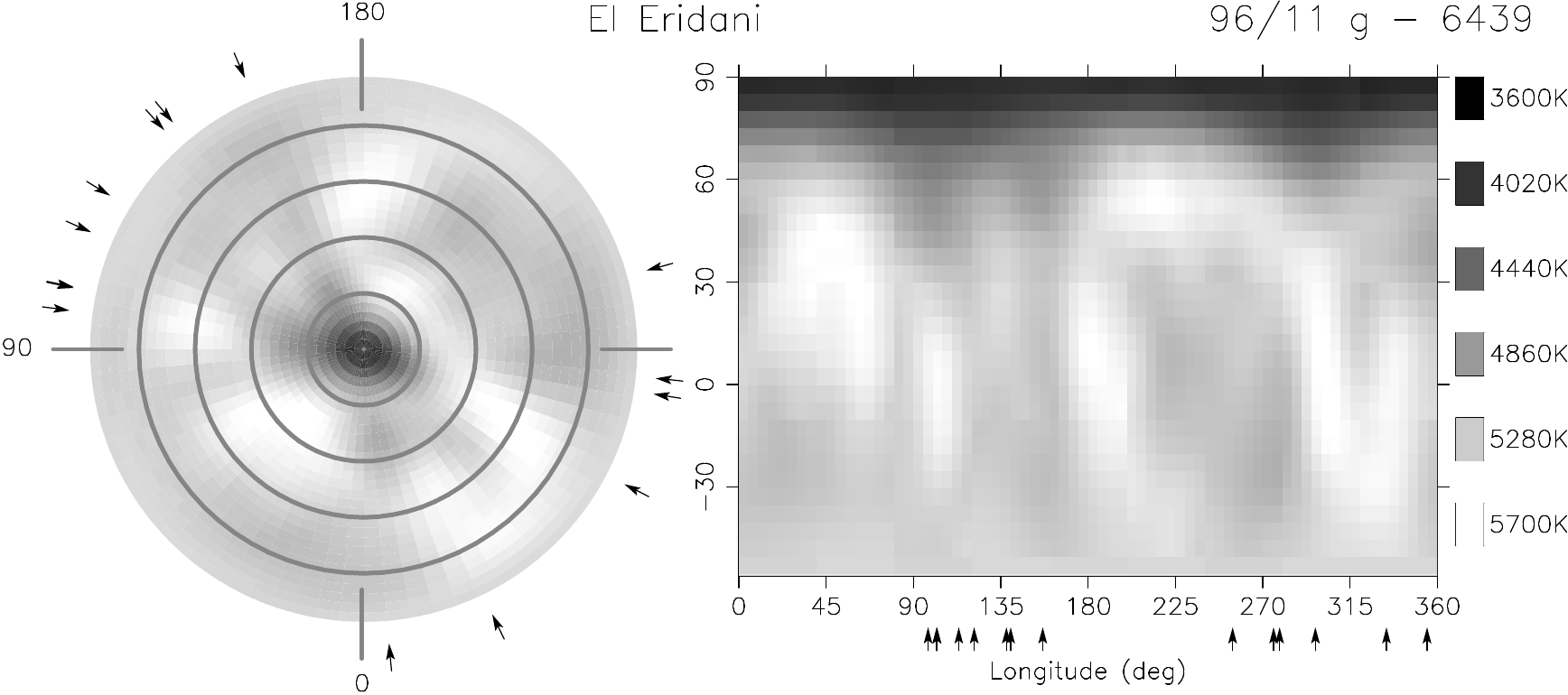}
\includegraphics[viewport=38 501 538 723,width=50mm,clip,angle=0]{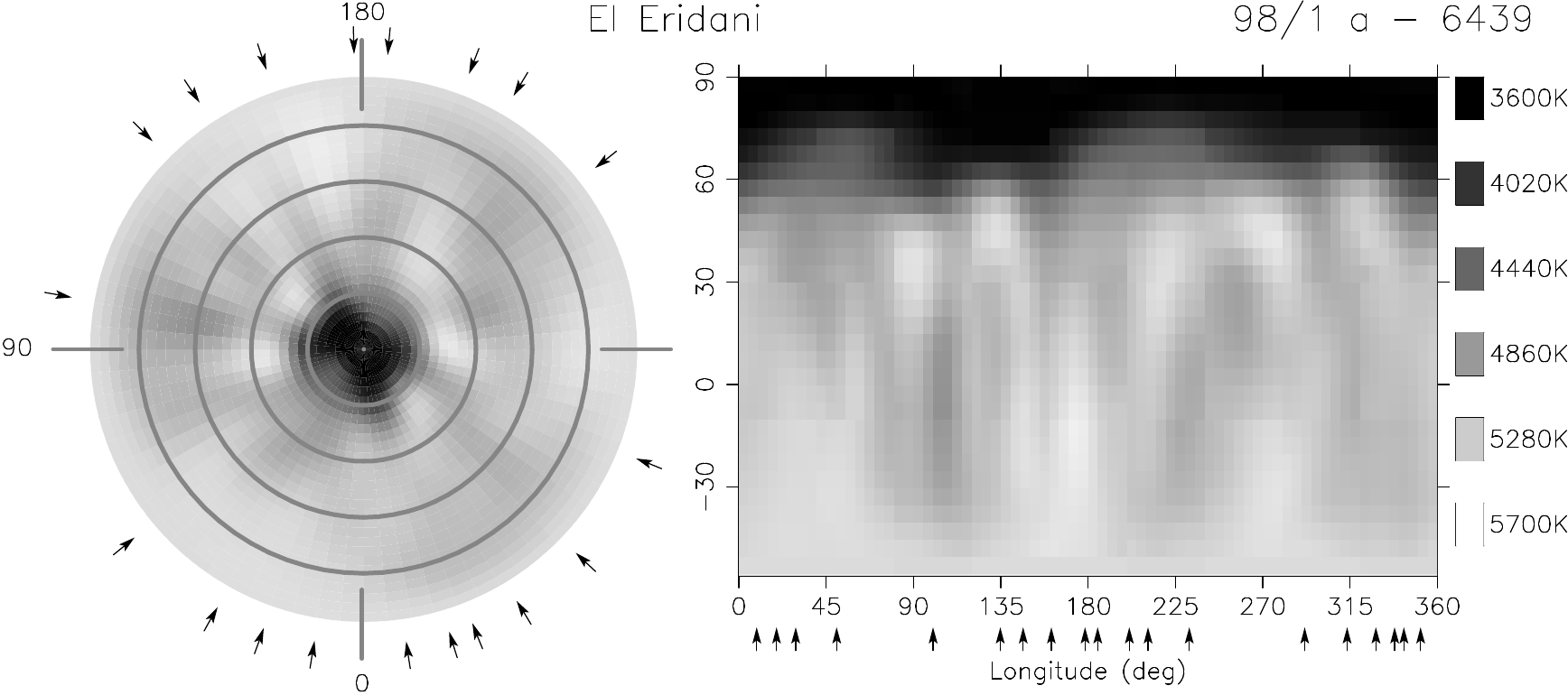}
\includegraphics[viewport=38 501 538 723,width=50mm,clip,angle=0]{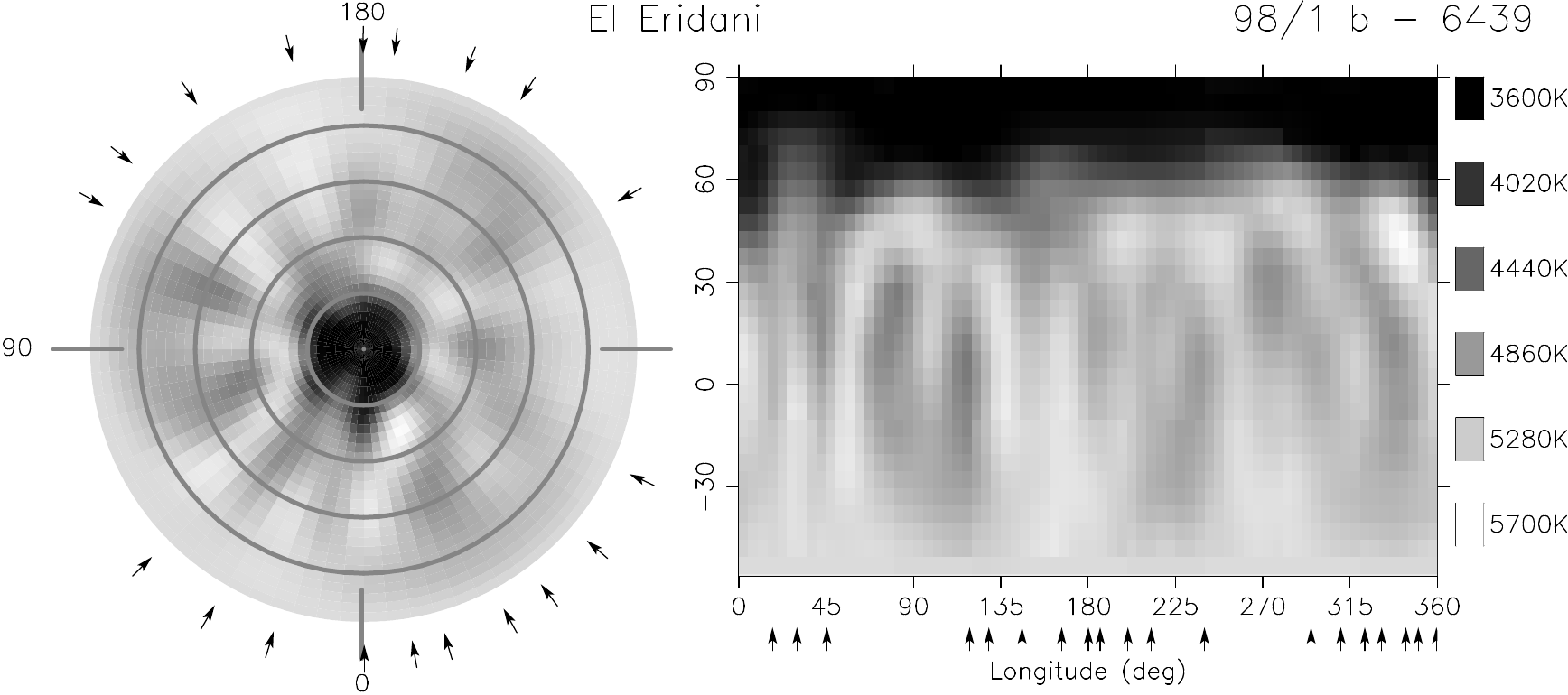}
\includegraphics[viewport=38 501 538 723,width=50mm,clip,angle=0]{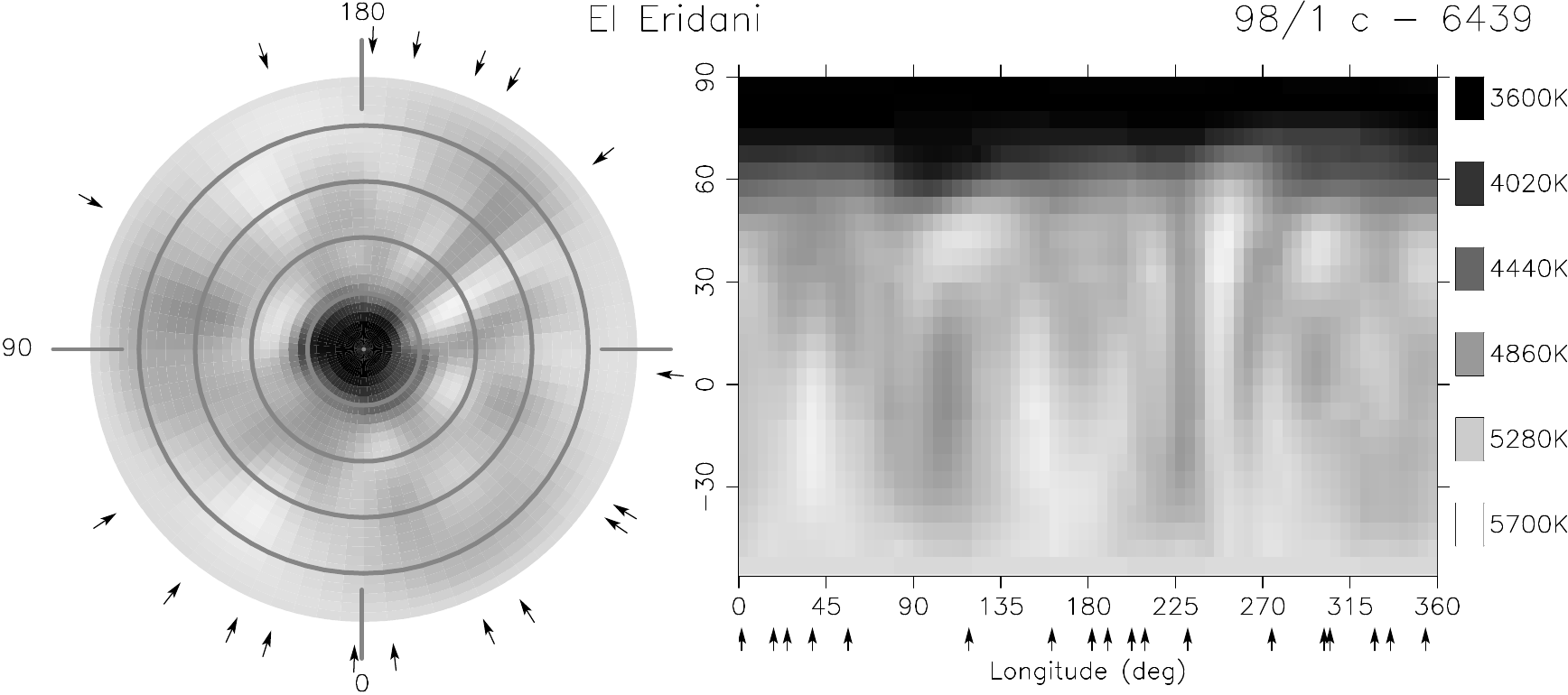}
\includegraphics[viewport=38 501 538 723,width=50mm,clip,angle=0]{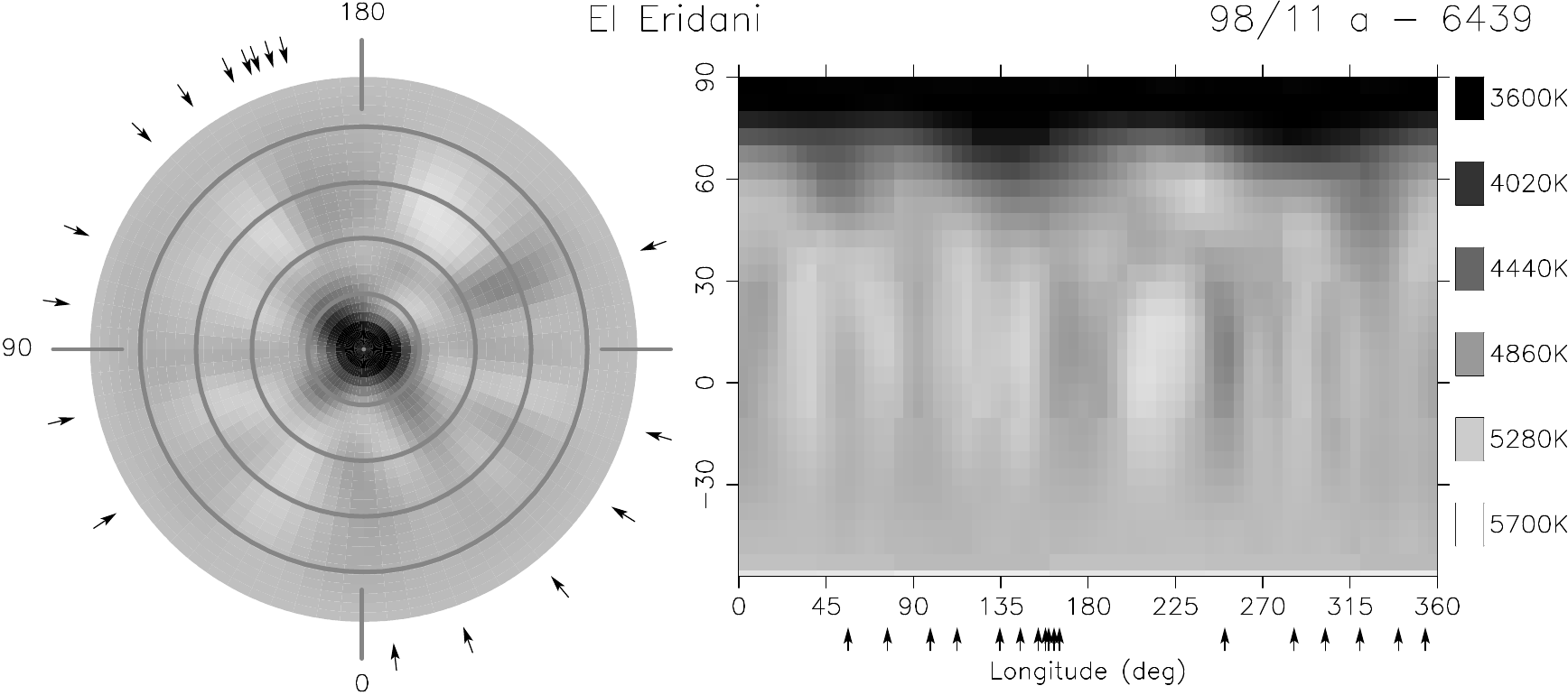}
\includegraphics[viewport=38 501 538 723,width=50mm,clip,angle=0]{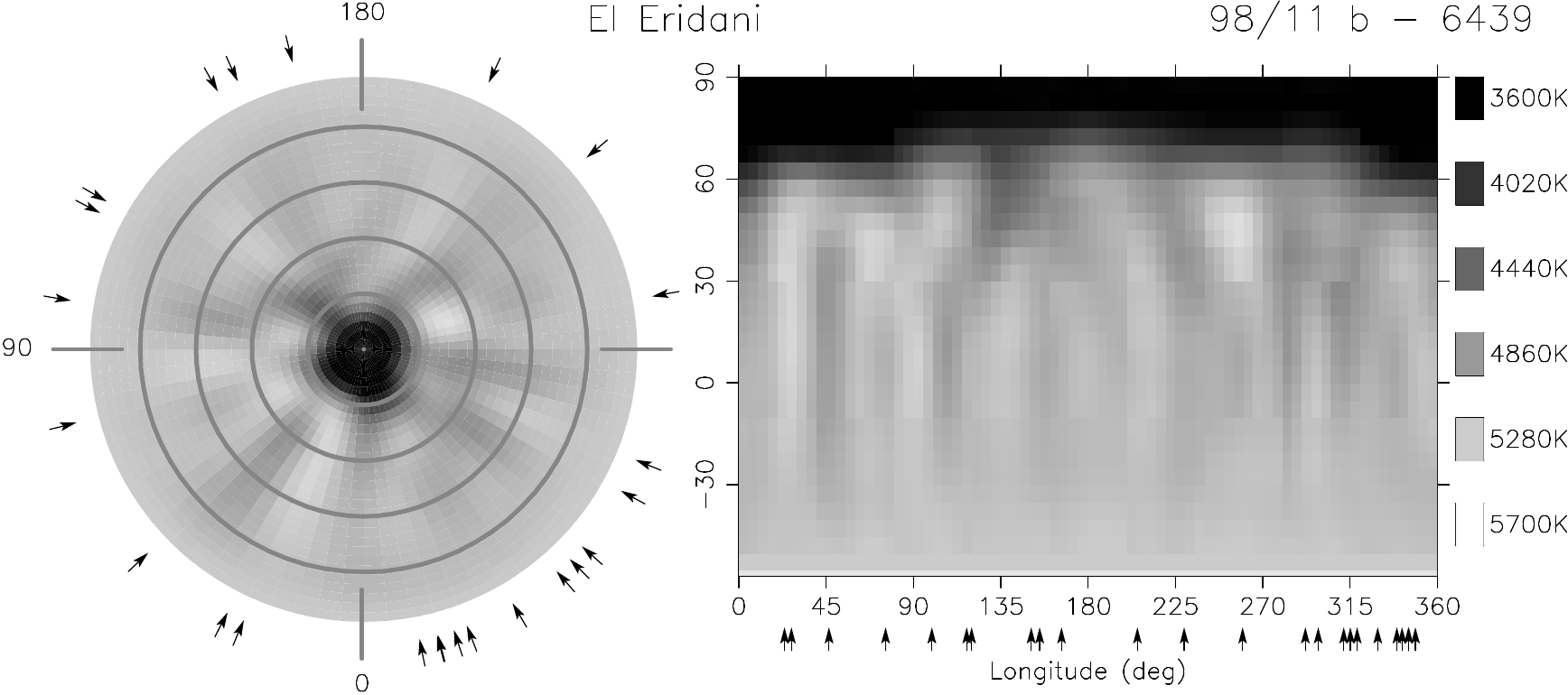}
\includegraphics[viewport=38 501 538 723,width=50mm,clip,angle=0]{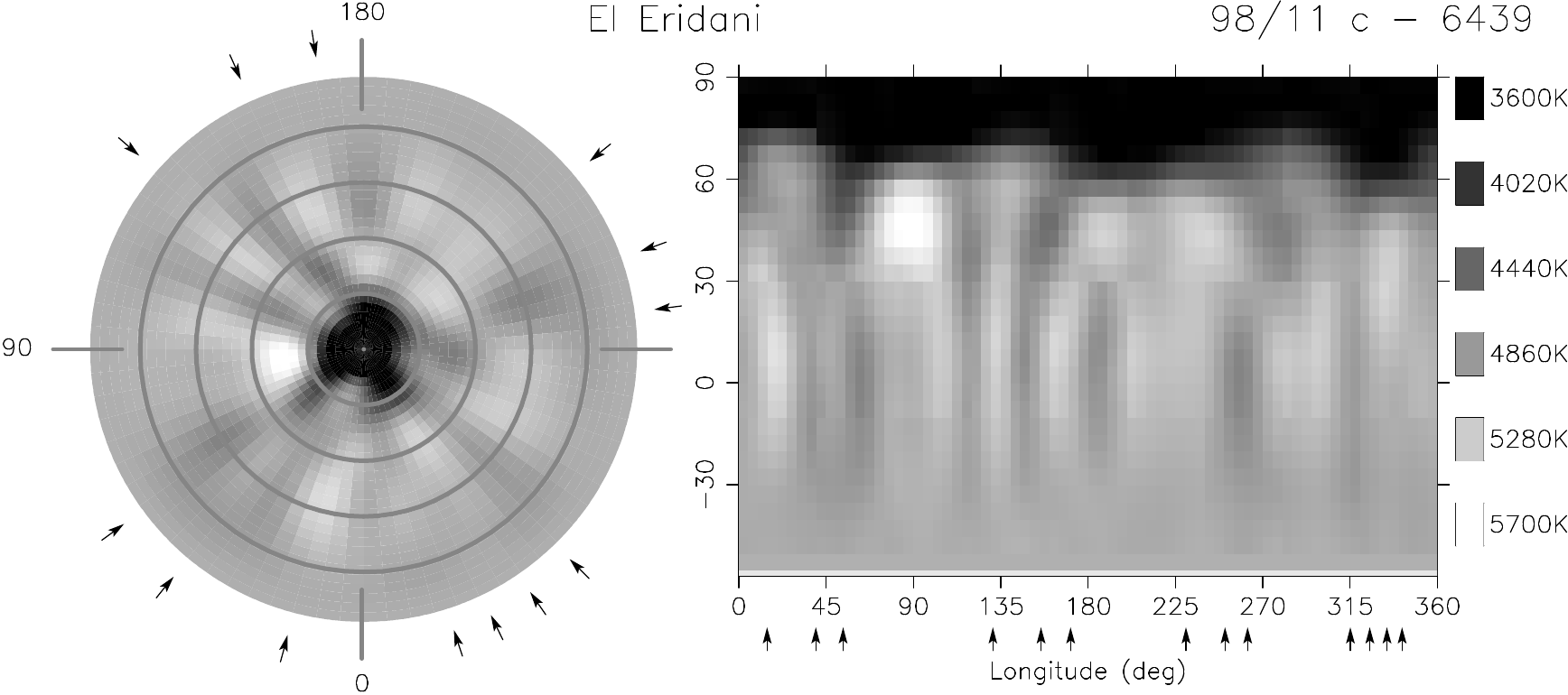}
\includegraphics[viewport=38 501 538 723,width=50mm,clip,angle=0]{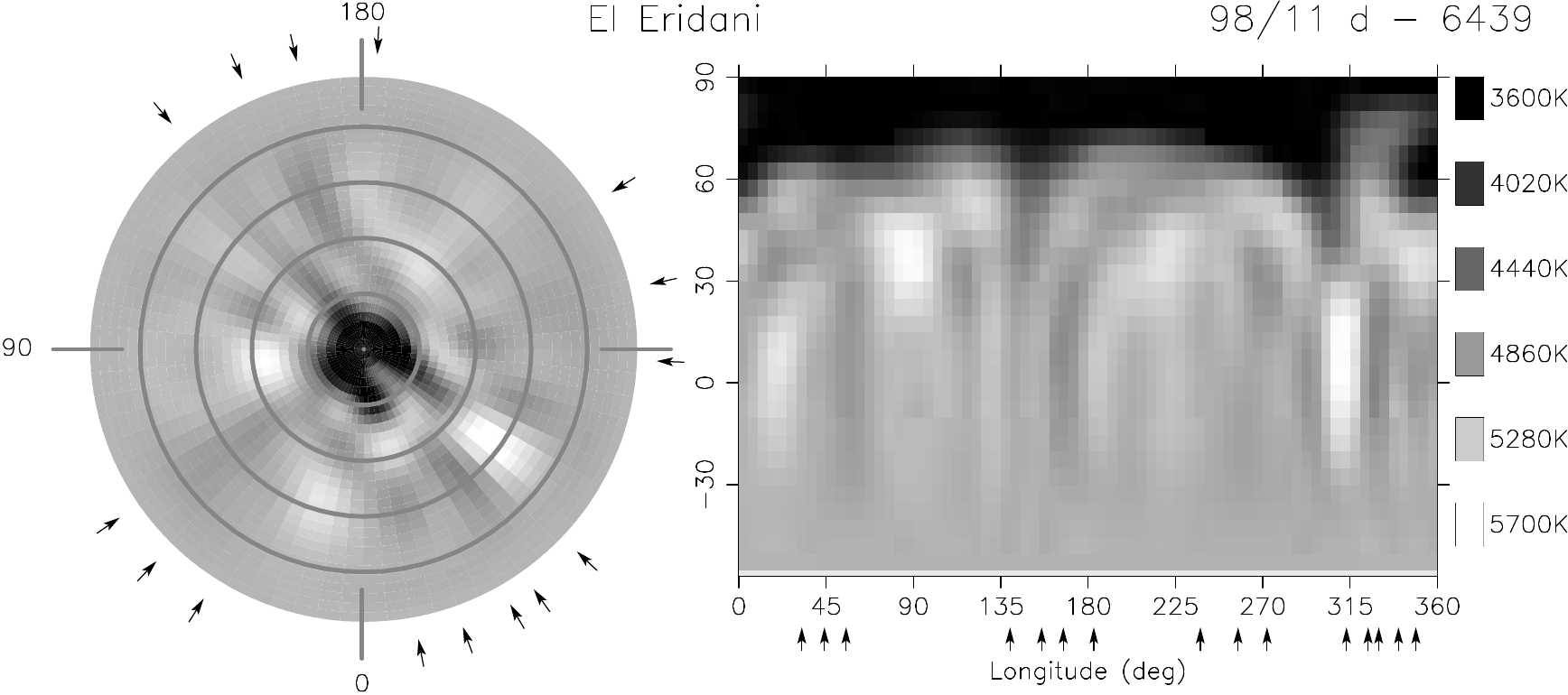}
\includegraphics[viewport=38 501 538 723,width=50mm,clip,angle=0]{fig/DI_B_empty.pdf}
\caption[]{
Comparison of all Doppler images from the Ca\,{\sc i} 6439
line, consisting of pole-on view (left part), with dark, solid circles
drawn at equally spaced intervals of 30\degr\ down to a latitude
of -30\degr ; and pseudo-Mercator projection (right part), from latitude
-56\degr\ to +90\degr . The phases of the observations are marked
by arrows around and below the maps, respectively.} \label{map}
\end{center}
\end{figure*}

\begin{table}
\begin{center}
\caption{Values used as input parameters for the Doppler imaging code {\tt TempMap}.}\label{tab:DIparam}
 \begin{tabular}{ll}
  \hline
  \hline
  \noalign{\smallskip}
  Parameter & Value     \\
  \noalign{\smallskip}
  \hline
  \noalign{\smallskip}
  $P_{\rm phot} = P_{\rm rot}$                  & 1.9472324    \\
  T$_{\rm 0, phot} = $T$_{\rm 0, rot}$          & 2448054.7109 \\
  $\gamma$                                      & 21.64 \kms   \\
  K1                                            & 26.83 \kms   \\
  $e$ (eccentricity)                            & 0 (adopted)  \\
  $T_{\rm phot}$                                & 5500 K       \\
  $T_{\rm max}$                                 & 6000 K       \\
  $v\sin i$                                     & 51.0 \kms    \\
  Inclination $i$                               & 56.0\degr    \\
  $\log g$                                      & 3.5          \\
  Micro turbulence $\xi$                        & 2.0 \kms     \\
  Macro turbulence $\zeta_{\rm r} = \zeta_{\rm t}$ & 4.0 \kms  \\
  Regularisation                                & Maximum Entropy \\
  Weight of phot. data                          & 0.1   \\
  $\log [Ca]$ abundance                                 & $-6.3$ (0.6\,dex below solar)\\
  $\log [Fe]$ abundance                                 & $-5.5$ (1.1\,dex below solar)\\
  \noalign{\smallskip}
  \hline
 \end{tabular}
\end{center}
\end{table}

For DI, it is important to obtain high signal-to-noise ratios
(S/N $\geq$ 200) at moderate to high spectral resolution
($\lambda$/$\Delta\lambda\ \approx$ 30--40\,000).
The wavelength region around 6420 \AA\ provides up to four
relatively unblended lines usable for DI: Ca\,I 6439, Fe\,I
6430, Fe\,I 6411, and Fe\,I 6393 and was therefore chosen for the single-order
spectrographs available at the McMath-Pierce and Coud\'e feed telescope.
An example spectrum is shown in Fig.~\ref{typical_spectrum}.

EI\,Eri's unfortunate rotational period of 1.947 days is
almost an integer multiple of the day/night cycle. As a consequence, one
ideally needs 20 nights of continuous observations from a single observing site
in order to achieve perfect phase coverage. In practice, 14 consecutive nights
are sufficient to give a good-quality Doppler image.
No correction for the contribution of the companion star had to be adopted
as EI\,Eri is a single-lined spectroscopic binary.
The time resolution of consecutive spectra was 2700 -- 3600\,s,
corresponding to a spot motion of 6~--~8$^{\circ}$ in longitude
on the central meridian of the stellar surface
or 5.2~--~6.9\,\kms\ (0.11~--~0.15\,\AA) in the spectral line profile.
This is still within one resolution element across the broadened line profile:
With a typical resolving power of 24\,000~--~36\,000 ($\lambda/\Delta\lambda$), we achieve
8~--~12 resolution elements across the stellar disk,
which corresponds to a velocity resolution of 8.3~--~12.5\,\kms\ (0.18~--~0.27\,\AA)
and a spatial resolution along the equator at the stellar meridian of approximately 10--14\degr.
The orbital smearing accounts to 2.7~--~3.6\,\kms\ (0.06~--~0.08\,\AA).
We therefore extended our integration limit to 60 minutes.

For the maps presented in this investigation, we apply the
Doppler-imaging code {\tt TempMap} by John Rice -- as described by
\citet{rice:wehlau89} and reviewed by \citet{piskunov:rice93},
\citet{rice96} and, most recently, by \citet{rice02}. All maps
were plotted using the same temperature scale from 3600 to
5500\,K.
The spot temperature was determined by \citet{kgs90} using standardised $V$ and
$R$ photometry and the Barnes-Evans relation \citep{barnes:evans78}. He gives a
temperature difference (star minus spot) of $1860 \pm 400$\,K and, with
$T_{\mbox{\scriptsize star}} = 5460$\,K,
yields an effective spot temperature of $3600 \pm 400$\,K.
\citet{oneal:saar96} derived spot temperatures by using the 7055 and
8860\,\AA\ bands of the titanium oxide molecule and determined, assuming a
photospheric value of $T_{\mbox{\scriptsize phot}} = 5600$\,K, a value
of $T_{\mbox{\scriptsize spot}} = 3700 \pm 200$\,K which was confirmed by \citet{oneal:neff98}.
In practice, the spot temperature is automatically chosen by {\tt TempMap}
and reaches down to 3600\,K.

All 30 independent maps for the $\lambda$6439 line are shown in Fig.~\ref{map}.
Like most maps, it resembles a steady pattern: an asymmetric polar spot
is escorted by several small low-latitude features. In some cases,
spots slightly warmer ($\approx$200 K) than the photosphere occur.
These are likely artifacts which are produced by the Doppler
imaging code due to, firstly, the use of differential photometry
instead of absolutely calibrated photometry and, secondly, due to
external uncertainties in the spectra \citep[see][]{rice:kgs00}.
Lower metallicity values lessen the appearance of artifical hot 
spots. Therefore, we assume the high divergence of the Fe abundance from 
the solar case of $-1.1$\,dex to be overdone and favour a more modest value of around 
$-0.5$\,dex. This value is in better agreement with the evolutionary 
tracks by \citet[ see paper I]{pietrinferni:cassisi04} and
is also supported by \citet{nordstrom:mayor04}.
However, recomputing the Fe Doppler maps with the more modest value of $-0.5$\,dex
does not alter the surface structure as such but only increases the temperature
of the artificial hot spots. Therefore, we used the lower metallicity value of $-1.1$\,dex
for computation of the Fe maps in order to improve their comparability.

\section{Analysis and parameterization of Doppler maps}
The spectroscopic data were inverted into a series of Doppler
images spanning 11 years and amounted to a total of about 100 images
for up to 28 independent epochs (34 images in $\lambda$6439, 30 in
$\lambda$6430 and 25 in $\lambda$6411 and a few in $\lambda$6393).
Mean separation in time between two independent maps is 141~days,
minimum and maximum separation is 3 (overlapping) and 382~days,
respectively.

\begin{figure}[tbh]
\begin{center}
\vspace{1cm}
\includegraphics[viewport=18 35 515 394,width=82mm,clip,angle=0]{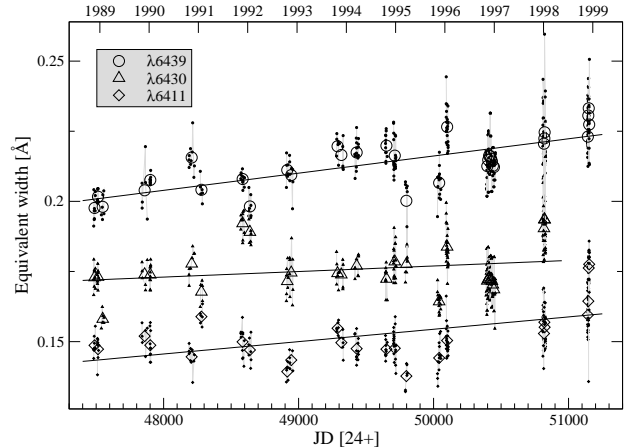}
\caption[Equivalent widths for all epochs] {Equivalent widths for
the three mapping lines at all epochs. The large open symbols are
the mean values for the data used in the respective Doppler maps,
the small symbols denote the individual spectra. The black lines
denote linear fits for each mapping line. The corresponding values
of the fits are listed in Tab.~\ref{tab:eqw}.}
\label{di:anal:eqw}
\end{center}
\end{figure}

\begin{figure}
\begin{center}
\includegraphics[viewport=13 26 430 361,width=82mm,clip,angle=0]{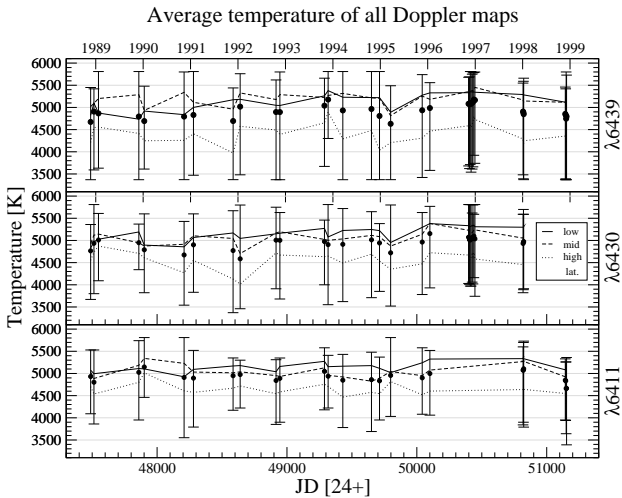}
\caption[Absolute temperature variations] {Temperature range for
all Doppler maps. Filled circles give the overall average of the
respective map, bars denote the minimum and maximum temperature.
The solid lines are the low-latitude averages (-55\degr\ --
0\degr), the dashed lines the mid-latitude averages (0\degr\ --
45\degr), and the dotted lines the high-latitude averages
(45\degr\ -- 90\degr).}\label{di:tempvar}
\end{center}
\end{figure}


Apart from tagging preferred spot locations, we extract several
parameters from the Doppler maps and carry out statistical
evaluations to see if there are any trends in the temperature
distribution or even correlations with the proposed activity cycle
\citep[see Fig.~2c in][ and Fig.~\ref{di:anal:phot} in this
paper]{olah:kgs02}.

\subsection{Line equivalent widths}

Fig.~\ref{di:anal:eqw} shows the equivalent widths (EW) for each
mapping line (small symbols) as well as the average EW (big symbols) for
each set of observations. A possible trend is the linear increase of the EW
during the course of the eleven years of observation.
This trend is seen with almost the same amount of 11.2 and 11.3\% (2.4 and
1.7$\sigma$) in the high-quality line $\lambda$6439 and the mid-quality
line $\lambda$6411, but only 4\% (0.8$\sigma$) in the poor-quality
$\lambda$6430 line (see Tab.~\ref{tab:eqw}).
If real, this could be suggestive of an increasing
chromospheric contribution from a magnetic cycle with a period
much longer than our observational coverage.

\begin{table}
\begin{center}
\caption{Increase of equivalent width, from Fig.~\ref{di:anal:eqw}.}\label{tab:eqw}
 \begin{tabular}{lccc}
  \hline
  \hline
  \noalign{\smallskip}
Wavelength		& $\lambda$6439 & $\lambda$6430 & $\lambda$6411 \\
  \noalign{\smallskip}
  \hline
  \noalign{\smallskip}
Initial value (fit) [\AA]	& 0.201		& 0.172		& 0.143 	\\
Final value (fit) [\AA]		& 0.223		& 0.179		& 0.159		\\
Increase value [\AA]		& 0.022		& 0.007		& 0.016		\\
Increase percentage		& 11.2\,\%	& 4.2\,\%	& 11.3\,\%	\\
Standard dev.			& 0.009		& 0.009		& 0.010		\\
Incr.\,value/Std.\,dev.		& 2.4$\sigma$	& 0.8$\sigma$	& 1.7$\sigma$	\\
  \noalign{\smallskip}
  \hline
 \end{tabular}
\end{center}

  \vspace{3mm}

\begin{center}
\caption{Increase of fractional spottedness, from Fig.~\ref{di:anal:fraction}.
Units are percentage points.}\label{tab:fraction}
 \begin{tabular}{lccc}
  \hline
  \hline
  \noalign{\smallskip}
Wavelength		& $\lambda$6439 & $\lambda$6430 & $\lambda$6411 \\
  \noalign{\smallskip}
  \hline
  \noalign{\smallskip}
Initial value (fit)	& 11.2		& 15.0		& 16.6 	\\
Final value (fit)	& 12.6		& 21.5		& 18.9	\\
Increase value]		& 1.47		& 6.43		& 2.24	\\
Increase percentage	& 13.2\,\%	& 42.7\,\%	& 13.5\,\%	\\
Standard dev.		& 2.7		& 3.4		& 1.8	\\
Incr.\,value/Std.\,dev.	& 0.54$\sigma$	& 1.9$\sigma$	& 1.2$\sigma$	\\
  \noalign{\smallskip}
  \hline
 \end{tabular}
\end{center}
\end{table}

\subsection{Surface temperature}

Fig.~\ref{di:tempvar} shows the progression of the average
temperature for all Doppler maps from 1989 to 1999, distinguished
for each mapping line\footnote {~Doppler maps of the $\lambda$6393
line are omitted in the following inspections as there are only
four maps of $\lambda$6393 available (one in 96/01 and three in
98/01).}. The filled circles are the overall temperature averages
of the corresponding Doppler maps, the bars in the axis of
ordinates denote the minimum and maximum temperature, the
respective upper solid line is the temperature average of the
low-latitude band, the dotted line that of the mid-latitude band
and the lower solid line that of the high-latitude band. The low,
mid, high-latitude bands run from $-$55\degr\ to 0\degr, from
0\degr\ to 45\degr, and from 45\degr\ to 90\degr,
respectively (smaller latitudes are not seen as they are tilted
out of sight due to the stellar inclination). Temperature variations
are noticed but not resembled in all spectral lines at the same time
and therefore are likely not real.
However, there seems to be a spot maximum (corresponding to
a minimum in temperature) in the high-latitude band in 1992 and
again in around 1995, seen in $\lambda$6439 and $\lambda$6430.
Overall, no pronounced correlation with the photometric variation
is evident though. Systematically wrong input values for the
line-profile computations can lead to spurious variations in the
overall temperature, e.g. caused by artificial hot spots. In order
to account for possible misfits, the latitude-dependent temperature
averages were normalized by the respective total average temperature.
More clearly, we now perceive the temperature decrease in high
latitudes in 1992 (in the $\lambda$6439 and $\lambda$6430 line)
which corresponds to an increase in the low-latitude band.
We also see a clear overall negative correlation between temperature
variations in the high and the low latitude bands. However, as
in the pre-normalized case, all variations seen are not resembled
in all spectral lines and therefore are likely not real. Above all,
we do not perceive an apparent correlation between temperature and
any other of the parameters derived from the Doppler maps.
In particular, no correlation with the long-term photometric
variations as seen in Fig.~\ref{di:anal:phot} and, more clearly,
in \citet[][ Fig.~2c]{olah:kgs02}, is evident.


\begin{figure}
\begin{center}
\includegraphics[viewport=30 357 552 587,width=80mm,clip,angle=0]{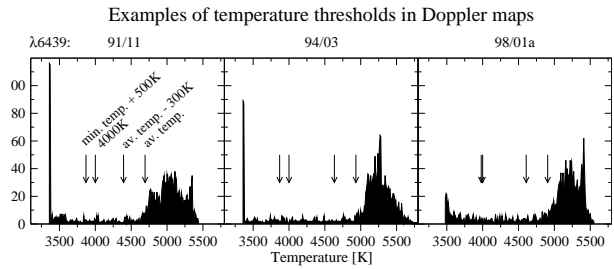}
\caption[Examples of temperature distributions in Doppler maps]
{Temperature threshold definitions for
three examples of temperature distributions in our Doppler maps
(Nov. 1993, Mar. 1994, Jan. 1998). The arrows represent several
threshold definitions for calculating the fractional spottedness
in Fig.~\ref{di:anal:fraction}. } \label{di:anal:tempdist}
\end{center}
\end{figure}

\begin{figure}
\begin{center}
\includegraphics[viewport=33 84 399 625,height=122mm,clip,angle=0]{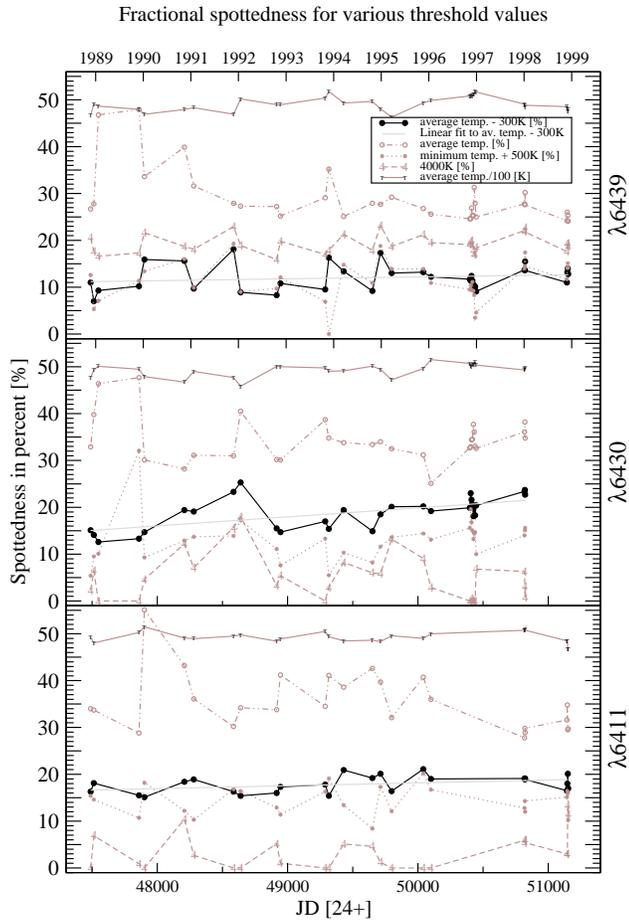}
\caption[Fractional spottedness as a function of time] {Fractional
spottedness in \% of the entire sphere for all Doppler maps as a function
of time and various threshhold values (explained in the insert).
The solid black line with filled circles, i.e. the average temperature minus
300\,K (solid black line), gives the most consistent results. The grey line denotes
the correponding linear fit, see Tab.~\ref{di:anal:fraction}. For comparison,
the average temperatur (divided by 100) is plotted.}
\label{di:anal:fraction}
\end{center}
\end{figure}

\begin{figure*}
\begin{center}
\includegraphics[viewport=25 32 599 256,width=170mm,clip,angle=0]{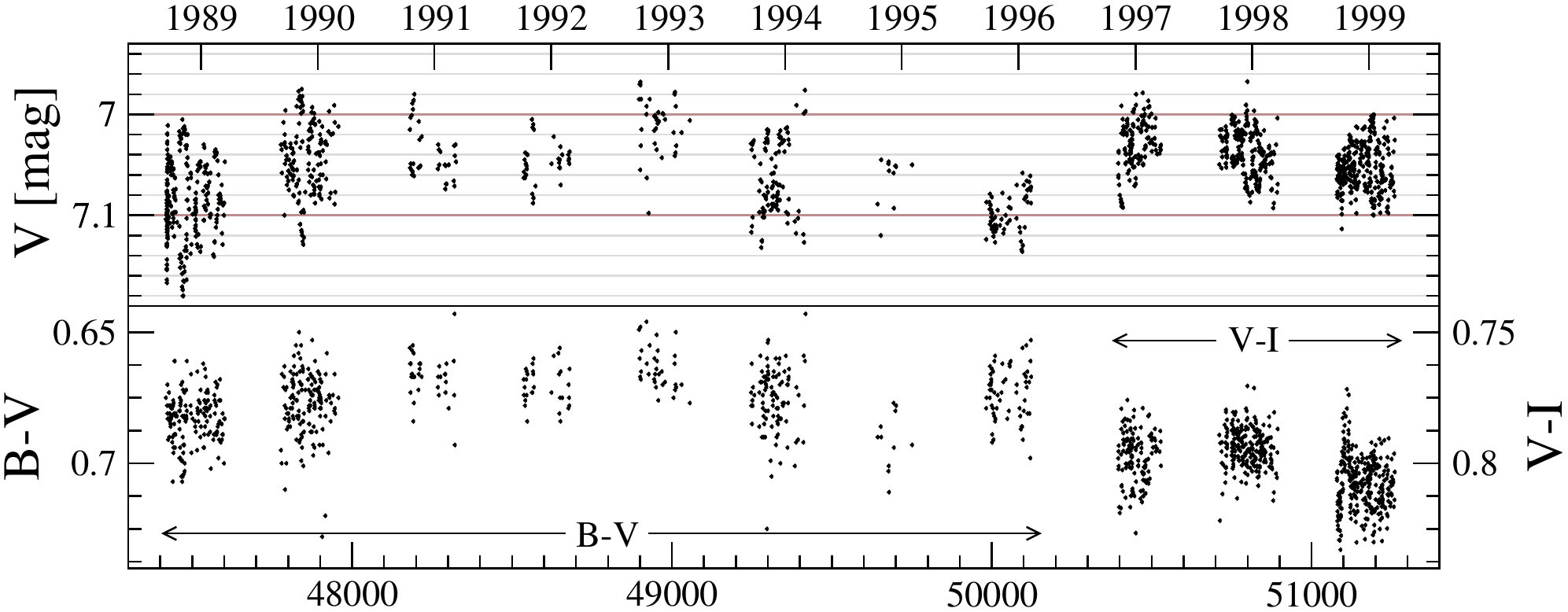}
\caption[Photometric data for the time coverage of our Doppler
maps.] {Photometric data for the time coverage of our Doppler
maps from \citet[][ for the complete data set, see their Fig.~2c]
{olah:kgs02}.}\label{di:anal:phot}
\end{center}
\end{figure*}

\subsection{Fractional spottedness}

We define a temperature threshold value and count the number
of pixels on the star's surface with temperatures below this
threshold and relate it to the total number of surface pixels.
Several options for defining a threshold are presented in
Fig.~\ref{di:anal:tempdist} which displays three examples of
temperature distributions
from the $\lambda$6439 line. First, the overall temperature average
is established, which usually sits at the lower edge of the
photospheric temperature domain (5000--5500K). Its location is, however,
so close to the edge that, as a second threshold option, a reduction of
the temperature average by a fixed value of 300\,K seems advisable.
Alternatively, we use the minimum value increased by 500\,K and
an arbitrarily fixed value of 4000\,K as threshold temperatures.
All temperature values smaller than the respective threshold
temperature provide us with a fraction value which is translated
to spottedness in percent. The results are shown in
Fig.~\ref{di:anal:fraction}. 
The threshold based on the average temperature is, as presumed, too
close to the photospheric-temperature bulge and sometimes exhibits
large increases (around 1990 which is not seen in any of the other
curves), induced by the scattering of the photospheric range.
The minimum
temperature, increased by a specific value, seems to be valuable
but sometimes exhibits large deviations which are not resembled in
any of the other mapping lines. Overall, we conclude that the
average temperature decreased by 300\,K gives the most reliable
threshold as it lies well between the photospheric and the
spot-temperature bulge.

Concentrating on the solid black line with filled circles
(Fig.~\ref{di:anal:fraction}), we notice an increase in spottedness in 
1992 of 10 percentage points in $\lambda$6439 (according to a factor 2),
also seen in the $\lambda$6430 map and another small
increase (4 to 6 percentage points) in 1995, mainly in the $\lambda$6439
map. We may compare these results with spot-filling factors derived from
TiO band modelling. \citet{solanki:unruh04} list filling factors
of EI\,Eri \citep[as obtained from][]{oneal:neff98} for the epochs
1992Mar (total 25\% / minimum 11\%), 1995Jan (13\%/5\%) and
1995Dec (13\%/5\%). These values coincide with the increase seen
in $\lambda$6430 in the map 1992Jan as compared to 1995Dec but not to
the same extent. However, this relation is not seen at all in
$\lambda$6439 and $\lambda$6411. Overall, no systematic changes
are evident and, above all, no correlation with the photometric
long-term trend is seen.

\begin{figure}
\begin{center}
\includegraphics[viewport=45 40 267 431,width=41.0mm,clip,angle=0]{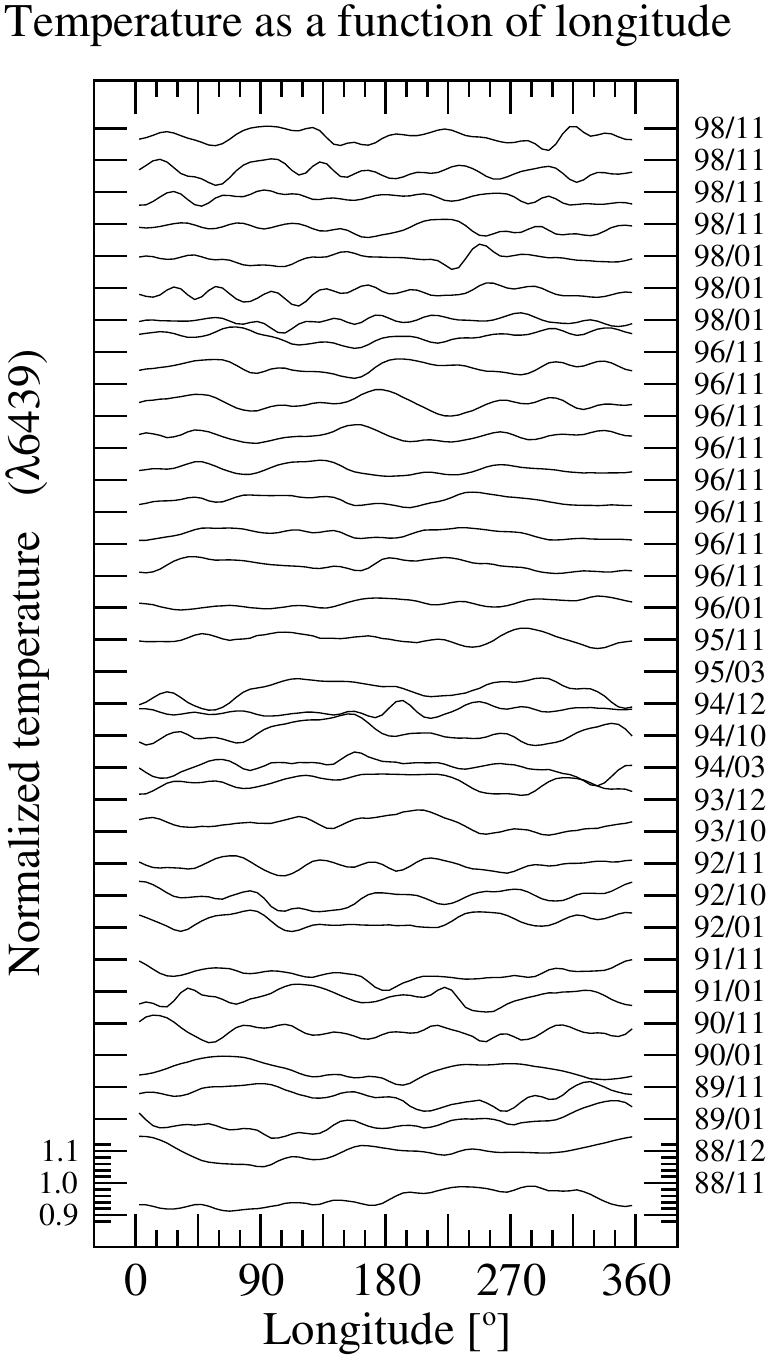}
\includegraphics[viewport=45 40 267 431,width=41.0mm,clip,angle=0]{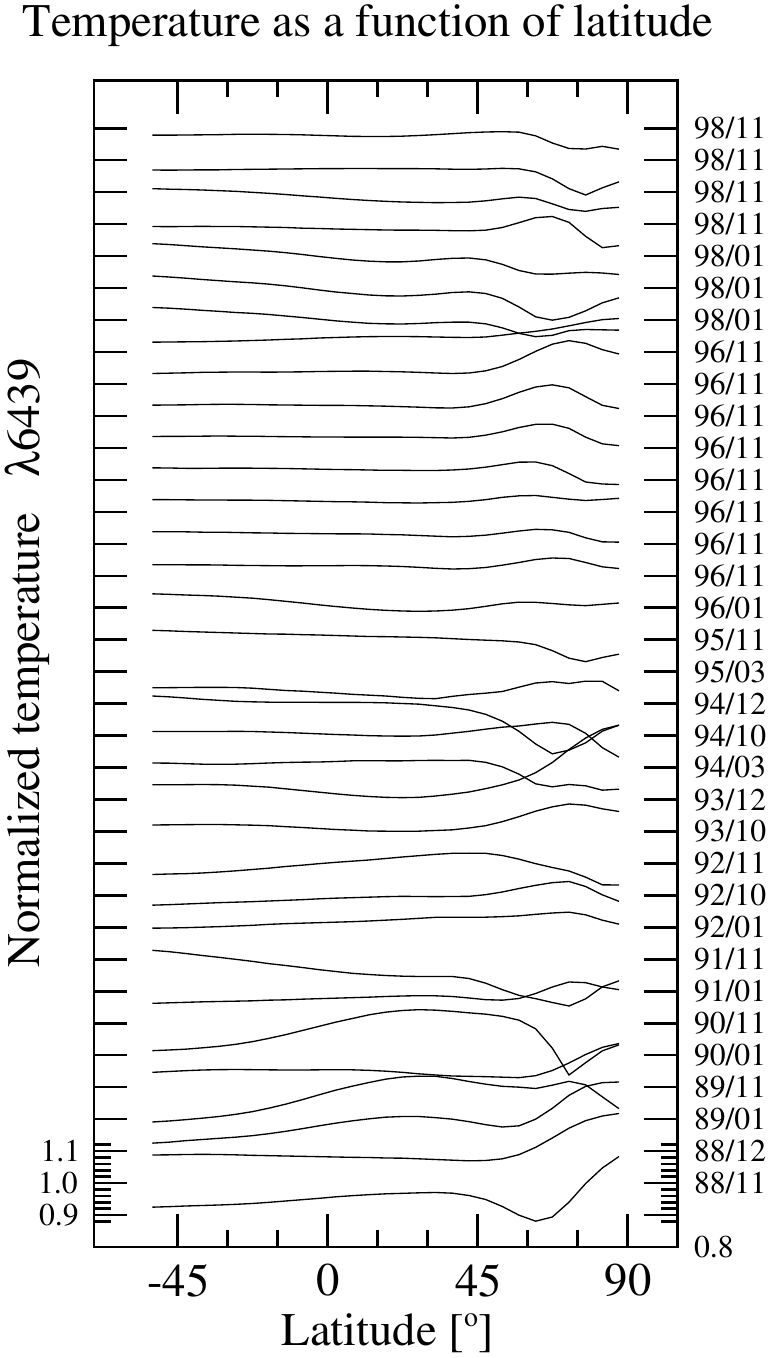}
\caption[Normalized temperature as a function of longitude/latitude]
{Temporal variations of surface temperature as a function of
longitude (left panel) and latitude (right panel). Each function is
normalized by the average of all functions and shifted by 0.1 for better 
viewing. To the right, year and month of observation is unscrambled.
The overall temperature profile shows only small and uniform variations.
The largest variations are found at high latitudes (65\,--\,80\degr),
as also seen in Fig.~\ref{di:grandaverage:sigma} (bottom panel).}
\label{di:anal:latlong}
\end{center}
\end{figure}

\begin{figure*}
\begin{center}
\includegraphics[viewport=50 346 609 500,width=170mm,clip,angle=0]{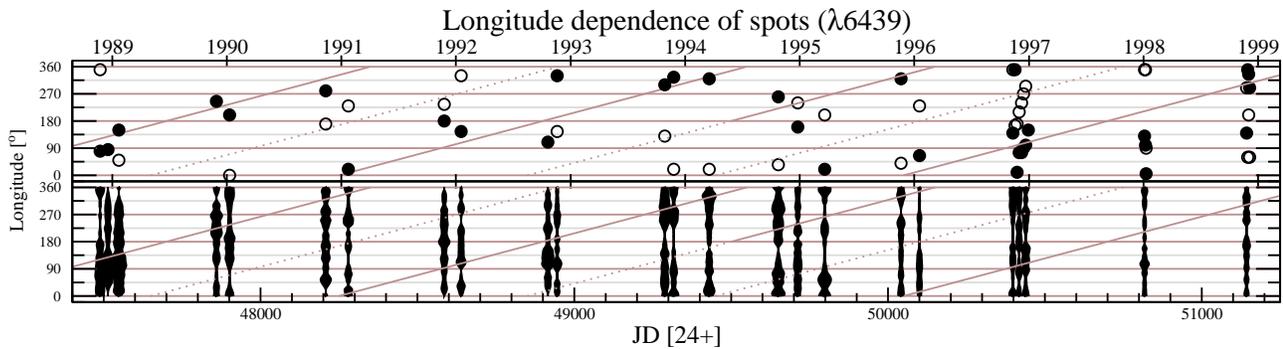}
\caption[Longitudinal spot-occurrence functions] {The upper
panel gives the longitude of the strongest (filled circles) and
second strongest (if existent; open circles) polar appendage for
each map. The tilted lines denote the suggested drifting active
longitudes. Primary active longitudes are shown as solid lines, 
secondary active longitudes (on the opposite side of the star)
as dotted lines. Lower panel: Longitudinal
spot-occurrence functions. The black areas represent
average temperature as a function of longitude, normalized by its
respective maximum average temperature. A broader area denotes
lower average temperature (corresponding to stronger spottedness)
at the respective longitude. Overlapping epochs were omitted.} \label{di:anal:long}
\end{center}
\end{figure*}


\begin{figure*}
\begin{center}
\includegraphics[viewport=50 386 586 481,width=170mm,clip,angle=0]{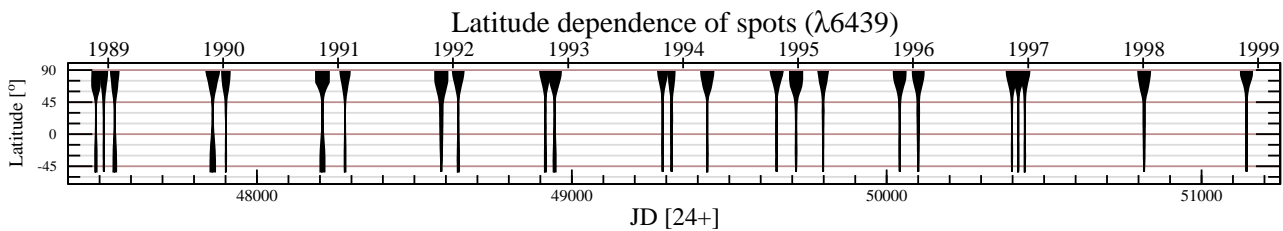}
\caption[Latitudinal spot-occurrence functions] {Latitudinal
spot-occurrence functions. The black areas mark averaged
temperatures as a function of latitude. See
Fig.~\ref{di:anal:long}b for further description. This display mode is
useful for comparing observational results with the theoretical
spot-probability functions presented by \citet{wooz2000}.}
\label{di:anal:lat}
\end{center}
\end{figure*}

\subsection{Spot occurence functions}

\begin{figure}
\begin{center}
\includegraphics[viewport=46 48 414 748,height=80mm,clip,angle=270]{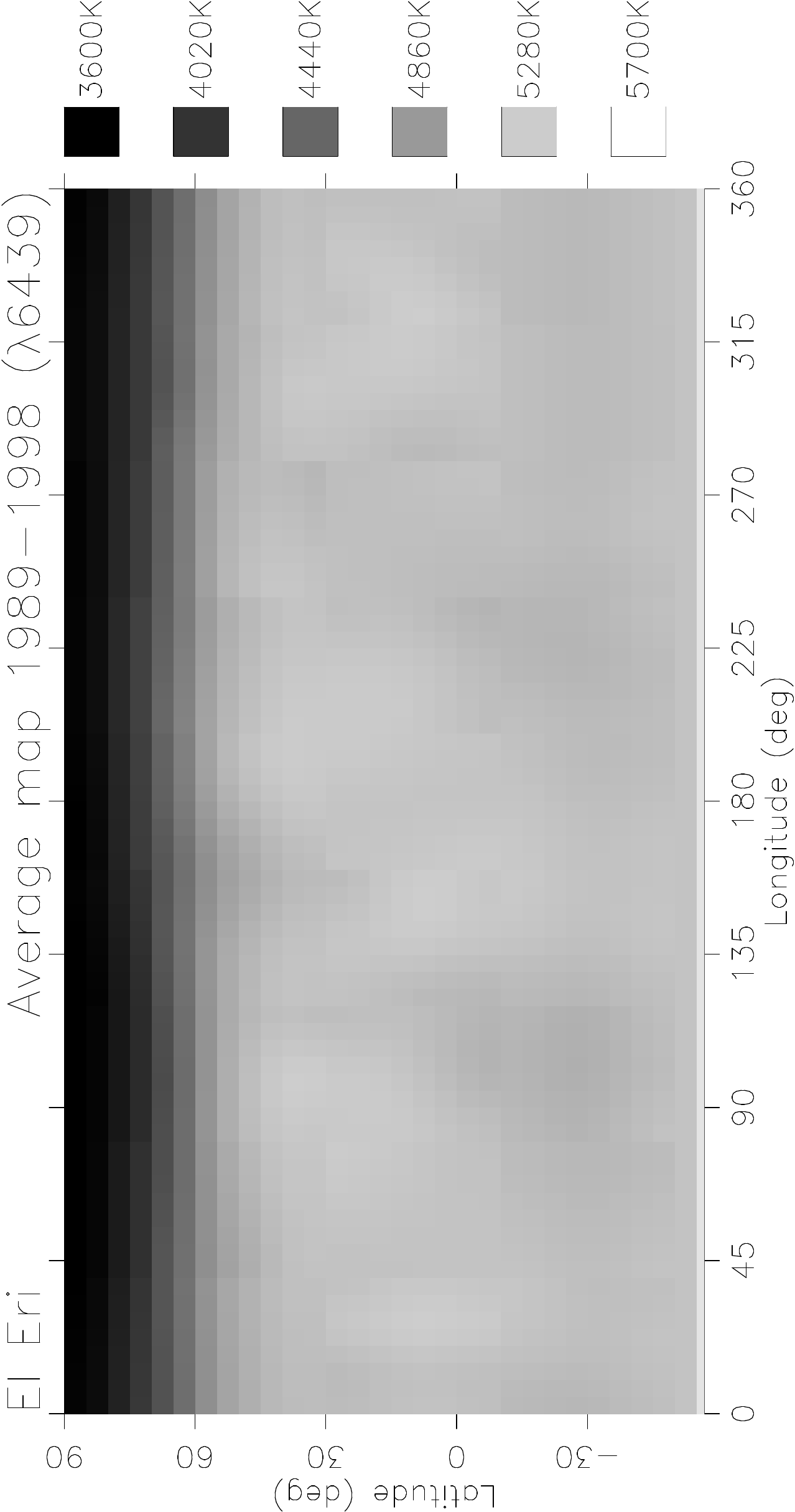}
\includegraphics[viewport=46 48 414 744,height=80mm,clip,angle=270]{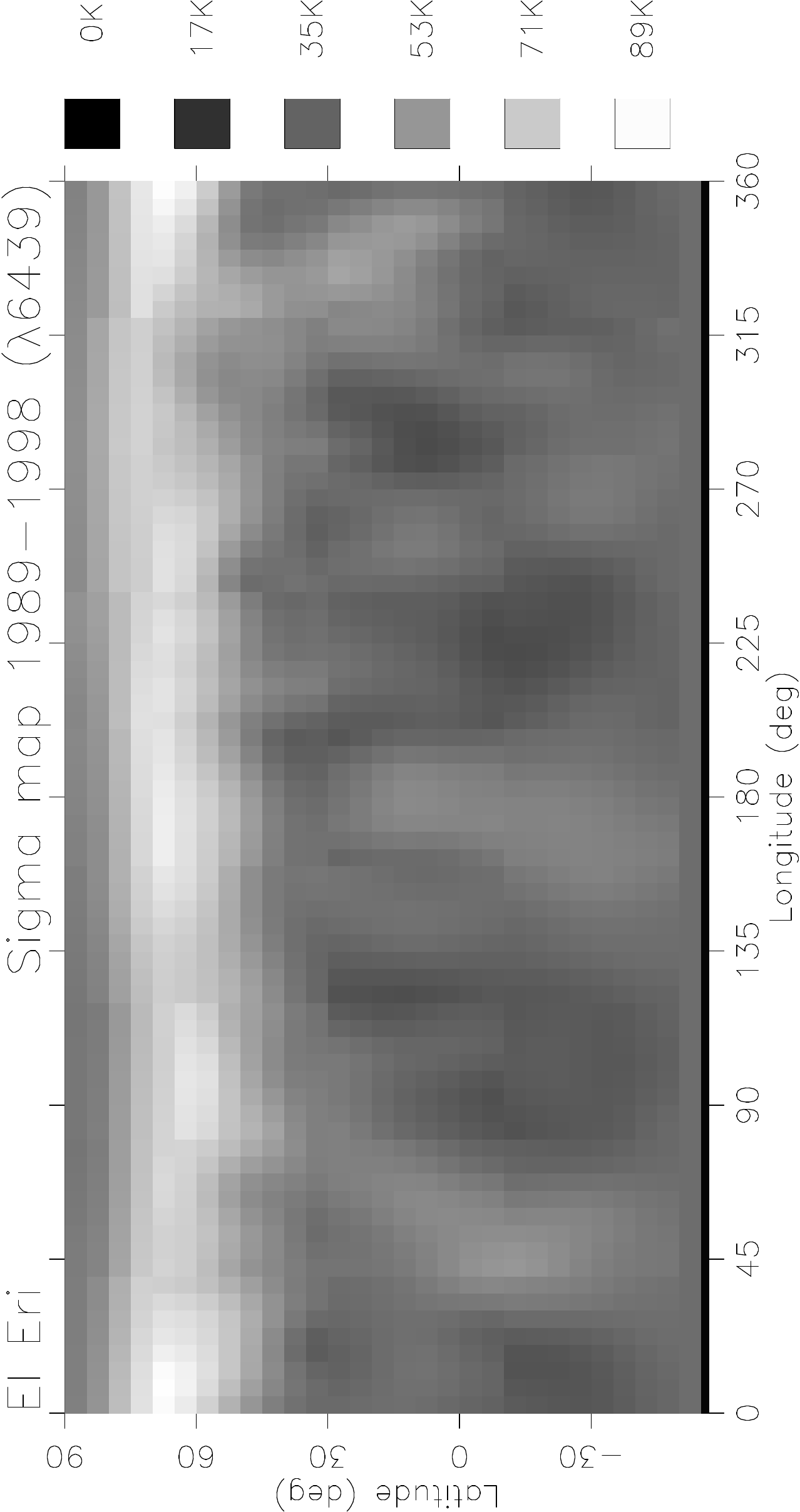}
\caption[``Grand average Doppler maps'' and sigma maps for
1988--1998] {Top: the grand average Doppler map for 1988--1998
($\lambda$6439). No preferred longitudes in the orbital reference
frame are seen. Bottom: Residual map for 1988--1998
($\lambda$6439).} \label{di:grandaverage:sigma}
\end{center}
\end{figure}

Fig.~\ref{di:anal:latlong} shows for all Doppler maps of the
$\lambda$6439 line the temperature distribution as a function of
stellar longitude (left) and latitude (right). Starting from
11/1988 on top, each map is shifted by $-$100\,K for better viewing. 
Overall, we cannot determine any preferred surface positions of spots
on EI\,Eridani (apart from the polar cap as such) which is also
demonstrated by the ``grand average Doppler map'', an average of
all available Doppler maps from 1988 through 1998
(Fig.~\ref{di:grandaverage:sigma}). Does this implicate that there
are no preferred longitudes at all? This would be in contradiction
to what was published by \citet{berdy:etal98b}. Their photometry
showed active longitudes which were not synchronized with the
orbital motion and were shifted by one orbital phase in about 2.7
years together with switching primary and secondary minima after
9~years (flip-flop). Fig.~\ref{di:anal:long} allows a more
detailed look at the spot occurrences as a function of stellar
longitude. The functions are the same as in
Fig.~\ref{di:anal:latlong} (left) but are plotted as two mirrored
contours filled in black, after normalization by their respective
maximum temperature (``spot-occurrence function''). A large width
denotes lower average temperature and therefore stronger
spottedness. The uppermost panel shows the latitudinal locations
of the strongest and the second strongest polar appendage as
inspected by eye. For the years 1989 till 1994,
\label{page:preflong} we see a possible correlation of the
longitudinal occurrences of polar appendages, showing a systematic
drift towards higher longitudes, i.e. later orbital phase,
indicated by solid (primary maxima) and dotted (secondary maxima)
grey lines. The shift amounts to 360\degr\ after about 3~years
which is very similar to the 2.7 years found by
\citet{berdy:etal98b}. Assuming a switch between primary and
secondary minima to occur around 1994 \citep[as proposed
by][]{berdy:etal98b}, we reverse the solid and dotted lines at
1994.4 in Fig.~\ref{di:anal:long} (uppermost panel) and see that
the continuation from 1994 onwards follows roughly this pattern.
The active longitudes noticed in Fig.~\ref{di:anal:long},
uppermost panel, are not clearly resembled by the
average-temperature functions shown in the lower panels of the
same figure (which give an average of each longitude, from $-i$ to
90\degr\ latitude). Thus, we conclude that the longitudinal
migration is mainly caused by the polar appendages. The
latitudinal region of the polar appendages is also the most active
area on EI\,Eri, as revealed by the long-term sigma maps (see
Fig.~\ref{di:grandaverage:sigma} bottom).

Fig.~\ref{di:anal:lat} shows the temperature averaged along
latitudes. Clearly, the polar cap dominates the surface maps.
Compare these results with the diagrams presented by \citet[][
Fig.~4]{wooz2000}. They are similar only to Granzer's 0.4\,\Msun\
cases, and no theoretically predicted spot-probability functions
are available yet for giant stars (the results from Granzer et al.
are for ZAMS stars).

The lower panel of Fig.~\ref{di:grandaverage:sigma} shows the
respective sigma map of a combination of all maps from 1988--1998
for the $\lambda$6439 line. Clearly, the most and strongest
variations are found in the appendices of the polar cap at a
latitude of 60\degr~--~75\degr. Nevertheless, as demonstrated by
Fig.~\ref{di:anal:lat}, the extent of the polar cap remains
very stable and ends, without exception, at $\approx$50\degr. The
polar cap is present in all maps.\footnote{Unlike wrongly claimed from
preliminary results by \citet{wasi:thinkshop} due to an ill-posed 
Doppler map and inconsistent DI parameters.} If the polar cap would
vanish for a certain epoch, we would expect the respective line
profiles to be more pointed, i.e. less shallow. While all averaged
profiles are very homogeneous, a few deviate slightly in line
depth. This behaviour is, however, not followed in all three
mapping lines simultaneously and is therefore likely due to
noise. In fact, none of the extracted parameters
exhibits a clear systematic variation in all three mapping lines.

\section{Discussion}
Our long-term DI study revealed that the brightness variations
(exhibiting a period of roughly 12 years) are not resembled by
any of the parameters
extracted from our Doppler maps nor by the directly measured line
equivalent widths. This is in contrast to the solar analog where
the total spottedness varies with the activity
cycle. Thus, we are asked to explain what exactly causes
the long-term variations of the mean brightness and why are these
variations not reflected in our Doppler maps.

\subsection{Short-term variations}
The photometric amplitude varies such that at the begin of the
photometric minima in 1984/85, as well as in 1995/96, the
amplitude was smallest and turned largest shortly thereafter. A
clear correlation with the long-term trend is not obvious
\citep[see Fig.~2c in][ left upper panel]{olah:kgs02}. These
short-term brightness variations are assumed to be caused by the
same spots moving across the visible disk of the star. However,
differences in the amplitude of those variations do not
necessarily reflect a larger or a smaller number of spots or
spottedness. They can also be a sign of a homogeneous or otherwise
a concentrated spot distribution on the star's surface. However,
the short-term brightness changes, if caused by spots, should be
directly related to the longitude functions in
Fig.~\ref{di:anal:long}, and should therefore be seen. Most
likely, the brightness changes are only caused by those parts of
the stellar surface that are not circumpolar. Therefore, we
recreated the plot displaying the longitudinal spot-occurrence
functions with the high latitudes (\,$90\degr - i$\,) excluded.
However, a relationship with the brightness variation 
is still not obvious.
After all, for epochs with a small number of photometric
data points, the disadvantageous rotational period of almost two
days may give an explanation: most photometric data of EI\,Eri
were collected by automatic telescopes which normally execute the
same targets every night. This brings about a similar time of
observation of a specific target for each night and, in the case
of EI\,Eridani, a similar orbital epoch which will, for a small
data sample, mimic smaller photometric amplitudes.

\subsection{Long-term variations}
A decrease or increase in total
brightness -- if caused by a larger spottedness or, like on the
Sun, by a higher faculae emission -- should accordingly be
reflected by the spottedness as extracted from the Doppler maps,
and thereby also by the average temperature and the equivalent
width. However, none of it is the case with EI\,Eri.
One possible explanation is that small inaccuracies in the
normalization of the spectra (continuum fitting) could potentially
influence the temperature span and thereby introduce noise.
Most likely though is that the cyclic brightness variation is caused
by a fluctuation of small unresolved spots while the fractional
coverage of large-scale spots remains basically constant. Doppler
images do a good job in catching starspots that significantly
modulate the line profile, but it is extremely difficult to detect
a background of small starspots evenly distributed over the
stellar surface. \citet{jeffers:aufdenberg06} obtained
spectrophotometric eclipse light curves with HST/STIS and mapped
the primary F9 component of the F9V+K4V-IV binary SV~Cam. They
found that the observed surface flux from the eclipsed
low-latitude regions of the F9 primary was 30\%\ lower than what
is predicted from model atmospheres and the \emph{Hipparcos}
parallax. This could be explained only if there is about a third
of the eclipsed region covered with unresolved cool starspots.
Even if this ``dark'' spot component is assumed to also exist over
the rest of the surface, a huge cap-like polar spot down to a
latitude of 48\degr\ was required in order to fit the eclipse
light curves. \citet{solanki:unruh04} found that even for heavily
spotted (hypothetical) stars, a large fraction of the spots are
smaller than the current resolution limit of Doppler images. Thus,
it is possible that a substantial part of the spot coverage is not
resembled by our measurements of fractional spottedness (as
presented in Fig.~\ref{di:anal:fraction}). Spot covering fractions
can also be derived from TiO band modelling, a method that does
not suffer from the resolution limitation. It would therefore be
essential to monitor and compare fractional spottedness as derived
from DI with TiO filling factors. However,
\citet{berdy02} obtained filling factors for the RS~CVn system
II~Peg that agree with the corresponding Doppler images without
requiring additional (unresolved) starspots. As a final
alternative, we could only assume that the long-term variation of
the total mean brightness is not caused by the degree of
spottedness. In this case, it would be questionable in what way
the photometric cycle corresponds to a magnetic cycle.

\subsection{Comparison with the Sun}
On the Sun, the solar cycle is visible not only in the visual
brightness \citep[see e.g.][]{willson:mordvinov03} but also in many
other tracers including total irradiance, radio and total magnetic
flux, \cahk\ emission, in magnetic features like sunspots, and others
\citep[see][]{harvey:white99}. The sunspot number, though, is the
most prominent one. Most indices vary roughly in phase with the
sunspot number. However, the modulation amplitude varies widely
between different indices. In visible light, the amplitude is
marginal and vague, and most likely not obvious to a casual 
observer.

The sunspot maximum correlates with a maximum in total
irradiance/brightness, \cahk core emission and total magnetic flux
\citep{harvey:white99,radick:lockwood98}. Thus, the Sun is
brighter during sunspot maximum which is caused by the dominating
emission from faculae. This behaviour is also found on other old
stars \citep{radick:lockwood98}. Both, the Sun and EI\,Eri, tend
to become bluer as they get brighter, 
which suggests that EI\,Eri might exhibit a higher level of
faculae emission during activity (i.e. photometric) maximum. On
the Sun, however, the emergence of faculae is usually associated
with spots.

A model presented by \citet{solanki:unruh98} indicates that the
radiative properties of magnetic features on the solar surface
(i.e. faculae and sunspots) provide the dominant contribution to
irradiance variations on a solar-cycle time-scale, 
suggesting that such features are also responsible for the visual
brightness variations on EI\,Eri. However, the total irradiance
variations seem to be strongly dependent on the inclination angle
of the star and, for the Sun, could increase by a factor exceeding
6 when viewed from a heliographic latitude of 60\degr
\citep[see][]{radick:lockwood98}.
Possibly, the variations in total spottedness on EI\,Eri are
minute and not resolved by our Doppler images while the
accompanying faculae could have an enhanced effect on the
brightness and photometric colour.

\subsection{Comparison with other stars}
The surprise that the proposed photometric cycle does not have a
corresponding spot or magnetic cycle might get support by results
from other stars. \citet{donati:acc03} presented long-term
magnetic surface images for the young K dwarfs AB~Doradus and
LQ~Hydrae and for the RS~CVn-type K1 subgiant HR\,1099 and found
that large regions with predominantly azimuthal magnetic fields
are continuously present at the surface of these stars. Long-term
structural changes take place but reflect no more than the limited
lifetime of the corresponding surface structures. An active
longitude is not confirmed on HR\,1099. Furthermore, no clear
secular change is detected in the axisymmetric component of the
magnetic field and in particular, no global polarity switch was
observed in the field ring pattern of HR\,1099, nor in the
high-latitude azimuthal-field ring of AB~Dor.
\citet{petit:donati04} report that, on HR\,1099, the small-scale
brightness and magnetic patterns undergo major changes within a
time-scale of 4--6 weeks, while the largest structures remain
stable over several years.

Another case, the \type{K2}{IV} RS~CVn star II~Peg, exhibits --
like EI\,Eri -- migrating active longitudes including the
flip-flop phenomenon. However, the total spot area is
approximately constant during this cycle. This means that II~Peg's
activity cycle is only expressed as the spot area evolution within
the active longitudes, i.e. as a rearrangement of the otherwise
nearly constant amount of the spot area \citep{berdy:etal99c}.

\citet{vogt:hatzes99}, having observed HR\,1099 for 11 years,
suggest starspots to be stellar analogs of solar coronal hole
structures and expect a dynamo cycle to be manifested by
periodicity in the area of the polar spot -- an expectation
not confirmed by EI~Eridani.

\subsection{Preferred longitudes}
As pointed out above, EI\,Eri seems to exhibit preferred
longitudes that migrate within the orbital reference frame,
supporting the finding from \citet{berdy:etal98b}. The time it
takes for an active longitude to shift by one orbital phase
(360\degr) is about 3~years which is close to the short-period
cycle found by \citet{olah:kgs02}. Can this shift be caused by
differential rotation? A evaluation of differential
rotation on EI\,Eri suggests a value of $\alpha \approx$ 0.037
\citep[i.e. solar, see the forthcoming paper][]{paper3}. If this value is real, the equatorial
spots overlap the polar region in about 50\,days. Particularly, differential
rotation then cannot account for the longitudinal migration of
spots as suggested by the phase drift of the migrating photometric
wave minimum -- a result also found on HR\,1099
\citep{vogt:hatzes99}. During the time the equator overlaps the
pole ($\approx$50\,days), the preferred latitudes shift only by
about 16\degr. After one orbital rotation, this shift amounts to
$\approx$\,0.65\degr, while the shift of the equatorial region comes to
about 14\degr\ during the same time. Therefore, considering both
differential rotation and the existence of stable,
preferred longitudes, the spots constituting the high-latitude
appendages must either be very short lived (otherwise they would
be moved out of the preferred longitudes by the slowerer rotating
polar region) or they are anchored in a deeper layer that is not
bound to the differentially rotating surface. Possibly, the
appendages at the preferred longitudes (or perhaps all spots in
the polar region) are the footprints of a strong dipole magnetic
field that is firmly anchored to the synchronized deeper core of
the star. The magnetic energy density of such a dipole could be
stronger than the kinetic energy density of the differential
shearing motions of the gas. This emergent dipole field would then
dominate the gas motions at high latitudes, producing apparent
synchronization of the high-latitude features with the orbit.

Active longitudes seem to be a common feature on active
close-binary stars. Such a phenomenon has been noticed using the
technique of Doppler imaging on, e.g., the active giant DM~UMa
\citep{hatzes95}, on the two main-sequence components of
$\sigma^2$~CrB \citep{kgs:rice03}, on the two pre-main sequence
components of V824 Ara \citep{hatzes:kurster99,paperxiv} and the
active giant UZ~Lib \citep{olah:kgs02}, as well as by long-term
photometric analysis on, e.g., RT~Lac \citep{cakirli:ibanoglu03},
UZ~Lib \citep{olah:kgs:wooz02uzlib} and on the RS~CVn stars
II~Peg, $\sigma$~Gem and HR~7275 \citep{berdy:etal98b}.
\citet{holzwarth:schuessler03b} presented a numerical
investigation for fast-rotating solar-like binaries which showed
that, although the magnitude of tidal effects is small, they
nevertheless lead to the formation of clusters of flux tube
eruptions at preferred longitudes on opposite hemispheres and
synchronized with the orbital motion. This finding is supported by
simplified calculations by \citet{moss:tuominen97} who showed that
synchronized close late-type binaries can be expected to exhibit
large-scale non-axisymmetric magnetic fields with maxima at the
longitudes corresponding to the two conjunctions (again locked
within the orbital reference frame). None of these investigations
can reproduce migrating active longitudes and the flip-flop
phenomenon as seen on EI\,Eri. For a possible theoretical
explanation of the flip-flop phenomenon on the single giant FK~Com
see \citet{elstner:korhonen05} \citep[see also][]{korhonen:elstner05,tuominen:berdy02}.

\section{Summary}

The key results from the Doppler-imaging analysis are as follows:

\begin{itemize}
\item The general morphology of the spot pattern persisted for more than 10 years.
\item The polar spot ($\Delta T\approx 1500$\,K) is stable.
No decay and reemergence due to, e.g., a polarity inversion was seen.
However, it changes its shape on short time-scales (one week).
The size of the polar cap is stable and reaches, without exception, down to
a latitude of $\approx$50\degr.
\item The most active surface region is the latitude between 60\degr\ and 75\degr;
the polar-spot appendages in this region seem to be responsible for the light
variability and especially for the migrating active longitudes.
\item Low latitude spots occur and decay on short time-scales (less than a week) and are
much less pronounced ($\Delta T\approx 500$\,K).
\item The polar appendages seem to appear on preferred longitudes
and drift towards larger longitudes with respect to the apsidal
line of the binary system with a period of $\approx$3\,yr, thereby
confirming the migrating active longitudes found earlier by
\citet{berdy:etal98b}. No fixed preferred longitudes locked to
the orbital reference frame of the system are seen though.
\item No correlation is seen between the photometric long-term cycle and any parameter extracted
from our spectra and Doppler maps; this includes equivalent width,
latitude and longitude function, average temperature and
spottedness.
\end{itemize}

\acknowledgements We are very grateful to the Deutsche
Forschungsgemeinschaft for grant STR 645/1 and for the
Hungarian-German Intergovernmental Grant D21/01. Special thanks go
to Thomas Granzer and J\'anos Bartus for their support and
computer assistance. This research project made extended use of
the SIMBAD database, operated at CDS, Strasbourg, France. The
Doppler imaging code {\sc TempMap} by John Rice was used for this
paper.



\bibliography{aa_mnem,kgs_etal,wasi}

\end{document}